\PassOptionsToPackage{unicode}{hyperref}
\PassOptionsToPackage{hyphens}{url}
\PassOptionsToPackage{dvipsnames,svgnames,x11names}{xcolor}
\documentclass[
  letterpaper,
  DIV=11,
  numbers=noendperiod]{scrartcl}
\usepackage{xcolor}
\usepackage{amsmath,amssymb}
\usepackage{iftex}
\ifPDFTeX
  \usepackage[T1]{fontenc}
  \usepackage[utf8]{inputenc}
  \usepackage{textcomp} % provide euro and other symbols
\else % if luatex or xetex
  \usepackage{unicode-math} % this also loads fontspec
  \defaultfontfeatures{Scale=MatchLowercase}
  \defaultfontfeatures[\rmfamily]{Ligatures=TeX,Scale=1}
\fi
\usepackage{lmodern}
\ifPDFTeX\else
\fi
\IfFileExists{upquote.sty}{\usepackage{upquote}}{}
\IfFileExists{microtype.sty}{% use microtype if available
  \usepackage[]{microtype}
  \UseMicrotypeSet[protrusion]{basicmath} % disable protrusion for tt fonts
}{}
\makeatletter
\@ifundefined{KOMAClassName}{% if non-KOMA class
  \IfFileExists{parskip.sty}{%
    \usepackage{parskip}
  }{% else
    \setlength{\parindent}{0pt}
    \setlength{\parskip}{6pt plus 2pt minus 1pt}}
}{% if KOMA class
  \KOMAoptions{parskip=half}}
\makeatother
\makeatletter
\ifx\paragraph\undefined\else
  \let\oldparagraph\paragraph
  \renewcommand{\paragraph}{
    \@ifstar
      \xxxParagraphStar
      \xxxParagraphNoStar
  }
  \newcommand{\xxxParagraphStar}[1]{\oldparagraph*{#1}\mbox{}}
  \newcommand{\xxxParagraphNoStar}[1]{\oldparagraph{#1}\mbox{}}
\fi
\ifx\subparagraph\undefined\else
  \let\oldsubparagraph\subparagraph
  \renewcommand{\subparagraph}{
    \@ifstar
      \xxxSubParagraphStar
      \xxxSubParagraphNoStar
  }
  \newcommand{\xxxSubParagraphStar}[1]{\oldsubparagraph*{#1}\mbox{}}
  \newcommand{\xxxSubParagraphNoStar}[1]{\oldsubparagraph{#1}\mbox{}}
\fi
\makeatother

\usepackage{color}
\usepackage{fancyvrb}

\DefineVerbatimEnvironment{Highlighting}{Verbatim}{commandchars=\\\{\}}
\usepackage{framed}
\definecolor{shadecolor}{RGB}{241,243,245}
\newenvironment{Shaded}{\begin{snugshade}}{\end{snugshade}}

\newcommand{\AttributeTok}[1]{\textcolor[rgb]{0.40,0.45,0.13}{#1}}

\newcommand{\CommentTok}[1]{\textcolor[rgb]{0.37,0.37,0.37}{#1}}

\newcommand{\ConstantTok}[1]{\textcolor[rgb]{0.56,0.35,0.01}{#1}}
\newcommand{\ControlFlowTok}[1]{\textcolor[rgb]{0.00,0.23,0.31}{\textbf{#1}}}

\newcommand{\DecValTok}[1]{\textcolor[rgb]{0.68,0.00,0.00}{#1}}
\newcommand{\DocumentationTok}[1]{\textcolor[rgb]{0.37,0.37,0.37}{\textit{#1}}}

\newcommand{\FloatTok}[1]{\textcolor[rgb]{0.68,0.00,0.00}{#1}}
\newcommand{\FunctionTok}[1]{\textcolor[rgb]{0.28,0.35,0.67}{#1}}

\newcommand{\NormalTok}[1]{\textcolor[rgb]{0.00,0.23,0.31}{#1}}

\newcommand{\OtherTok}[1]{\textcolor[rgb]{0.00,0.23,0.31}{#1}}

\newcommand{\SpecialCharTok}[1]{\textcolor[rgb]{0.37,0.37,0.37}{#1}}

\newcommand{\StringTok}[1]{\textcolor[rgb]{0.13,0.47,0.30}{#1}}

\usepackage{longtable,booktabs,array}
\usepackage{calc} % for calculating minipage widths
\usepackage{etoolbox}
\makeatletter
\patchcmd\longtable{\par}{\if@noskipsec\mbox{}\fi\par}{}{}
\makeatother
\IfFileExists{footnotehyper.sty}{\usepackage{footnotehyper}}{\usepackage{footnote}}
\makesavenoteenv{longtable}
\usepackage{graphicx}
\makeatletter
\newsavebox\pandoc@box
\newcommand*\pandocbounded[1]{% scales image to fit in text height/width
  \sbox\pandoc@box{#1}%
  \Gscale@div\@tempa{\textheight}{\dimexpr\ht\pandoc@box+\dp\pandoc@box\relax}%
  \Gscale@div\@tempb{\linewidth}{\wd\pandoc@box}%
  \ifdim\@tempb\p@<\@tempa\p@\let\@tempa\@tempb\fi% select the smaller of both
  \ifdim\@tempa\p@<\p@\scalebox{\@tempa}{\usebox\pandoc@box}%
  \else\usebox{\pandoc@box}%
  \fi%
}
\def\fps@figure{htbp}
\makeatother

\NewDocumentCommand\citeproctext{}{}

\makeatletter
 \let\@cite@ofmt\@firstofone
 \def\@biblabel#1{}
 \def\@cite#1#2{{#1\if@tempswa , #2\fi}}
\makeatother
\newlength{\cslhangindent}
\newlength{\csllabelwidth}
\newenvironment{CSLReferences}[2] % #1 hanging-indent, #2 entry-spacing
 {\begin{list}{}{%
  \setlength{\itemindent}{0pt}
  \setlength{\leftmargin}{0pt}
  \setlength{\parsep}{0pt}
  \ifodd #1
   \setlength{\leftmargin}{\cslhangindent}
   \setlength{\itemindent}{-1\cslhangindent}
  \fi
  \setlength{\itemsep}{#2\baselineskip}}}
 {\end{list}}
\usepackage{calc}

\newcommand{\CSLLeftMargin}[1]{\parbox[t]{\csllabelwidth}{\strut#1\strut}}
\newcommand{\CSLRightInline}[1]{\parbox[t]{\linewidth - \csllabelwidth}{\strut#1\strut}}

\KOMAoption{captions}{tableheading}
\usepackage{float}
\usepackage{eso-pic}
\makeatletter
\@ifpackageloaded{caption}{}{\usepackage{caption}}
\AtBeginDocument{%
\ifdefined\contentsname
  \renewcommand*\contentsname{Table of contents}
\else
  \newcommand\contentsname{Table of contents}
\fi
\ifdefined\listfigurename
  \renewcommand*\listfigurename{List of Figures}
\else
  \newcommand\listfigurename{List of Figures}
\fi
\ifdefined\listtablename
  \renewcommand*\listtablename{List of Tables}
\else
  \newcommand\listtablename{List of Tables}
\fi
\ifdefined\figurename
  \renewcommand*\figurename{Figure}
\else
  \newcommand\figurename{Figure}
\fi
\ifdefined\tablename
  \renewcommand*\tablename{Table}
\else
  \newcommand\tablename{Table}
\fi
}
\@ifpackageloaded{float}{}{\usepackage{float}}
\floatstyle{ruled}
\@ifundefined{c@chapter}{\newfloat{codelisting}{h}{lop}}{\newfloat{codelisting}{h}{lop}[chapter]}
\floatname{codelisting}{Listing}

\makeatother
\makeatletter
\@ifpackageloaded{caption}{}{\usepackage{caption}}
\@ifpackageloaded{subcaption}{}{\usepackage{subcaption}}
\makeatother
\usepackage{bookmark}
\IfFileExists{xurl.sty}{\usepackage{xurl}}{} % add URL line breaks if available
\makeatletter
\@ifundefined{xmpquote}{}{}
\makeatother
\hypersetup{
  pdftitle={Simple{,} general strategies for correcting the upward bias of calibration summaries of prediction models},
  pdfauthor={Mohsen Sadatsafavi1,*, Ramin Rezaeianzadeh1, Hessam Sadatsafavi2},
  colorlinks=true,
  linkcolor={blue},
  filecolor={Maroon},
  citecolor={Blue},
  urlcolor={Blue},
  pdfcreator={LaTeX via pandoc}}

\title{Simple, general strategies for correcting the upward bias of
calibration summaries of prediction models}
\author{Mohsen Sadatsafavi\textsuperscript{1,*}, Ramin
Rezaeianzadeh\textsuperscript{1}, Hessam Sadatsafavi\textsuperscript{2}}
\date{}
\begin{document}
\maketitle
\begin{abstract}
For clinical prediction models, calibration across the spectrum of
predicted values is often assessed via flexible calibration curves and
associated scalar summaries such as the integrated calibration index
(ICI). However, these scalar summaries tend to have a finite-sample
upward bias, as they are based on the absolute value of an estimate
whose sampling distribution can cross zero. We show the mechanism
underlying this bias and quantify its magnitude as a function of the
true calibration error and its variance. This mechanism motivates easily
implementable methods for reducing the bias (plug-in and inversion
methods, each applied locally or globally) that are applicable to any
smoothing method that generates an estimate and its sampling variance
for each point on the calibration curve. These methods are general in
that 1) they can be used for different outcome types (continuous,
binary, time-to-event, as well as counterfactual outcomes for models
that predict treatment benefit), and 2) the resulting bias-adjusted
calibration error curves yield bias-corrected estimates of other metrics
such as quantile-based ones (e.g., E50 and E90). Focusing on ICI as an
exemplary metric, numerical and case studies demonstrate that the upward
bias can be substantial, especially when the sample is small and the
true ICI is close to zero. Results show that the Global Inversion method
removed the most bias, particularly when the bias was substantial,
although it could slightly increase the mean absolute error if the bias
was small.
\end{abstract}

\begingroup
\renewcommand\thefootnote{}
\footnotetext{\textsuperscript{1}Respiratory Evaluation Sciences Program, Faculty of Pharmaceutical Sciences, The University of British Columbia, Vancouver, British Columbia, Canada. \textsuperscript{2}Gillings School of Global Public Health, University of North Carolina, Chapel Hill, North Carolina, USA. \textsuperscript{*}Corresponding author: Mohsen Sadatsafavi,
mohsen.sadatsafavi@ubc.ca.}
\endgroup

\clearpage

\section{Introduction}\label{introduction}

Calibration, how close the outputs of a predictive algorithm are to
their actual counterparts, is an important component of the performance
of clinical prediction models. Models that are miscalibrated can mislead
patients about the prospect of their outcome, and might cause providers
to make incorrect management decisions for the patient. The assessment
of model calibration is an integral part of external validation of a
model before its adoption for clinical use {[}1--3{]}. While calibration
has mostly been discussed for models that predict outcome probabilities,
the concept is extendable to other paradigms such as models that predict
individualized treatment effects (ITEs) {[}4{]}.

Calibration can be assessed in multiple ways {[}5{]}. Of particular
relevance is studying how close actual and predicted values are among
the strata defined by predicted values (e.g., if the outcome occurs in
X\% of patients whose predicted outcome probability is X\%). This type
of calibration has been referred to as `weak' or `moderate' calibration
by different authors (henceforth we refer to it as moderate calibration,
following the calibration hierarchy of Van Calster et al.) {[}1,6,7{]}.
As most models generate continuous (or near continuous) predictions,
computing the conditional mean for a given point requires regularization
(binning or smoothing) {[}8{]}, with common smoothing methods being
locally estimated scatterplot smoothing (LOESS), restricted cubic
splines (RCS), and generalized additive models (GAM) {[}9{]}. The
results are often visualized as a flexible calibration curve, which can
be accompanied by pointwise confidence bands to reflect the uncertainty
of the estimated curve.

When comparing the performance of competing models in a sample, or the
same model across different samples (e.g., across subgroups as part of
fairness assessment), summarizing the calibration curve into a scalar
metric of miscalibration can facilitate objective comparisons. Several
such metrics have been introduced; prominent among them is the
integrated calibration index (ICI) {[}9{]}. ICI is the average absolute
calibration error across the entire range of predicted risks.
Geometrically, it can be interpreted as the absolute area between the
flexible calibration curve and the line of perfect calibration, weighted
by the density of predicted values at each point {[}9{]}. In machine
learning, the analogous metric is the expected calibration error (ECE),
traditionally computed by binning predictions {[}10,11{]} but more
recently by kernel smoothing {[}12{]}. Quantile-based metrics such as
E50 (median of absolute calibration errors) and E90 (90th percentile of
such errors) are also common {[}9{]}. While these metrics have generally
been discussed for risk prediction models, they can be extended to
models that predict continuous outcomes {[}2{]}, or models that generate
counterfactual predictions for individualized treatment effects
{[}4,13{]}.

Importantly, these summary metrics tend to suffer from an upward bias.
Indeed, if the prediction model is fully calibrated, the true value of
ICI will be zero, but its estimate will always be positive, as the
calibration curve inevitably moves away from the line of perfect
calibration along its path. Compared to clinical prediction modeling,
bias in the summary indices of calibration is more widely discussed in
other fields, in particular machine learning and meteorology
{[}14--17{]}. Much of the attention has been given to measures of
quadratic, not absolute, error {[}14--16{]}. To the best of our
knowledge, mitigating the bias of ICI and quantile-based metrics
associated with a flexible calibration curve has not been sufficiently
explored, with the closest attempt being a recent cross-validated
estimator only applicable to ICI {[}18{]}.

This is an important gap, as a biased estimate of ICI and quantile-based
metrics can affect the overall verdict on model calibration. Calibration
is a criterion for model fairness {[}19--21{]}, and reporting guidelines
recommend that model performance be reported for key subgroups {[}3{]}.
As we will show, the bias of summary metrics might be substantially
different across subgroups of different sizes. This can result in
misleading judgments about the fairness of the model, with apparent
miscalibration being exaggerated in small (e.g., under-represented)
groups.

Here, we propose general approaches for correcting the finite-sample
bias of ICI and quantile-based metrics. Our bias correction strategies
are easily implementable, and are general on two fronts: 1) they can be
applied to any smoother that generates an estimate of local calibration
error and its variance; 2) as they only require the output of a fitted
curve, they can be directly applied to any outcome type, including
binary, ordinal, censored time-to-event, or counterfactual (e.g., models
that predict treatment benefit) outcomes, in a manner compatible with
the method used to construct the calibration curve.

\section{Methods}\label{methods}

We focus on ICI for a binary outcome from a previously developed model
now being assessed (validated) in an independent sample of size \(n\) of
\emph{iid} individuals. While we primarily focus on ICI, we study the
extension of this methodology to quantile-based metrics such as E50 and
E90. The focus on a binary outcome is mainly for brevity, as the
proposed bias correction strategies are generally applicable to any
outcome for which a flexible calibration curve can be constructed.

By \(\hat{P}(\boldsymbol X)\) and \(Y\) we denote the predicted value
and observed response of an individual with predictor values
\(\boldsymbol X\) from the target population, and by
\(e(\boldsymbol X) = g\{\hat{P}(\boldsymbol X)\} - \hat{P}(\boldsymbol X)\),
where \(g(p) = E(Y|\hat{P}=p)\), the true calibration error at that
predicted value. The subscript \(i\) denotes these quantities for the
\(i\)th individual in the sample. The true ICI is the expected absolute
calibration error,

\[ ICI = E|e(\boldsymbol X)|. \]

If the true calibration errors were known, their sample mean,
\(\frac{1}{n}\sum_{i=1}^n |e_i|\), would be an unbiased estimator of
ICI. However, the true calibration errors are not observed. Smoothing
methods such as LOESS, RCS, or GAM estimate this calibration error by
\(\hat{e}_i = \hat{E}(Y | \hat{P}=\hat{P}_i) - \hat{P}_i\), alongside a
standard error (SE) of the estimate (\(\hat{\sigma}_i\)). The estimated
ICI is \[\widehat{ICI} = \frac{1}{n}\sum_{i=1}^n |\hat{e}_i|.\] The SE
is often used to construct 95\% confidence bands around the curve, which
are typically \(\pm 1.96\) SE, justified by the Central Limit Theorem
{[}22{]}.

Unfortunately, even if individual \(\hat{e}\)s are unbiased estimators
of the true \(e\)s, their absolute values, \(|\hat{e}|\), are generally
biased estimators of \(|e|\), as explained below.

\subsection{The mechanism: the folding bias}\label{sec-folding}

\begin{figure}[H]
\centering
\caption{Illustration of folding bias. Left: a hypothetical true calibration curve
(black), with the shading showing the sampling distribution of curves estimated
from repeated validation samples. Right, for each point: the top row shows the sampling distribution of the smoothed calibration error $\hat{e}_i$; the bottom row shows the distribution of its absolute value $|\hat{e}_i|$.}
\label{fig:folding}
\includegraphics[width=\textwidth]{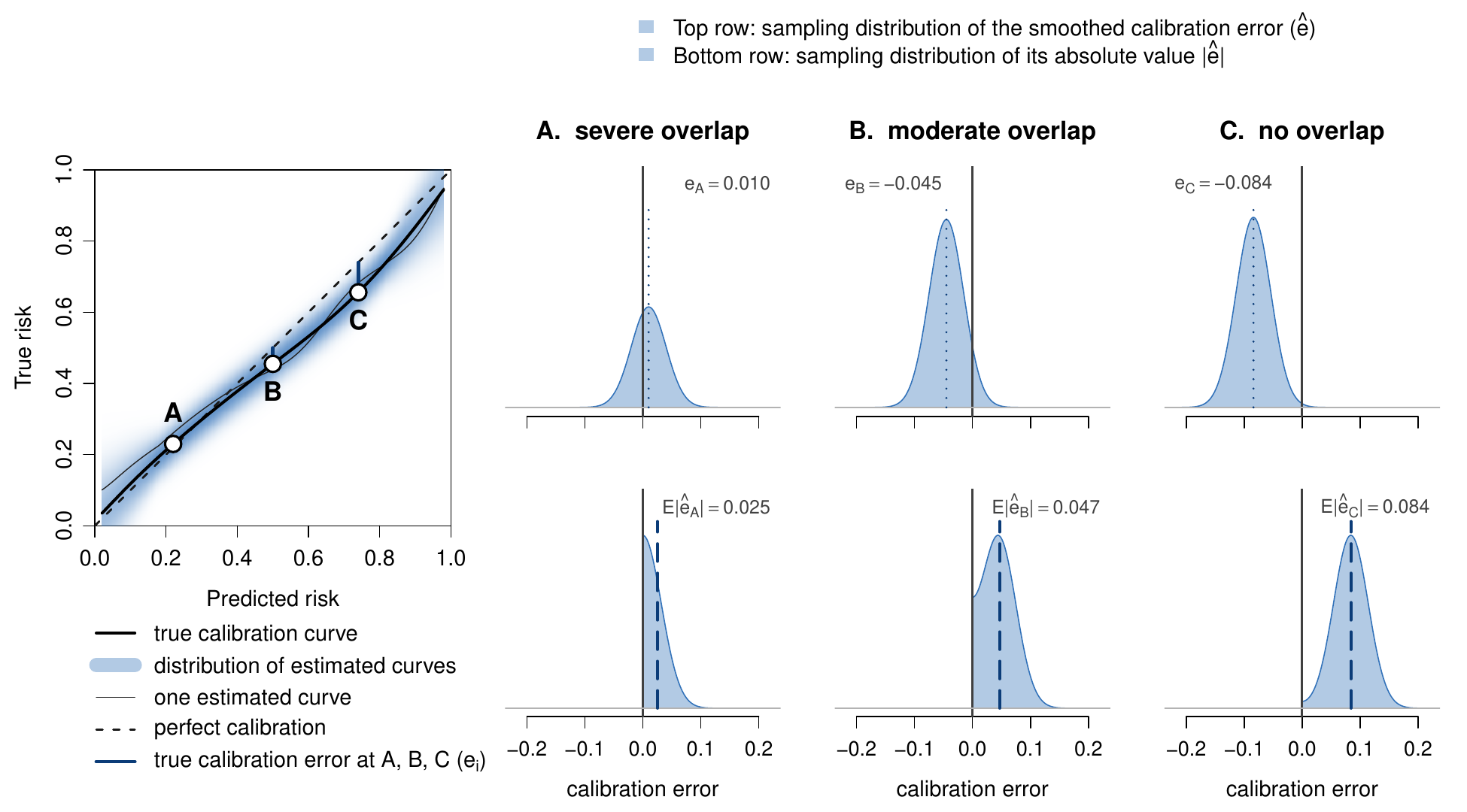}
\end{figure}

To explain the mechanism of the upward bias, which also informs our bias
correction strategies, we focus on Figure \ref{fig:folding}. This figure
depicts a hypothetical true calibration curve, together with the
distribution of curves that would be estimated from repeated validation
samples, and three exemplary points on the curve. The three dotted dark
blue vertical lines in the top row show the values of the actual
calibration errors \(e_i\) and the densities in the top row represent
the sampling distributions of the estimated error \(\hat{e}_i\).
Consider point A of this figure, where the true calibration error is
0.010 (and thus the absolute calibration error is 0.010). Even if the
smoothing algorithm itself is unbiased, that is, \(E(\hat{e}_A)=0.010\),
the \emph{expected} absolute value of the estimated calibration error,
\(E(|\hat{e}_A|)\), is 0.025, larger than the true absolute calibration
error. This is because the sampling distribution of calibration error
crosses zero, and thus when taking its absolute value, it `folds' around
zero.

The degree of the bias varies point by point. This bias happens only
when the sampling distribution has some density on both sides of zero,
and its extent depends on both the extent of the overlap and how far the
distribution stretches away from 0. At point B in Figure
\ref{fig:folding}, where the sampling distribution of calibration error
is further away from zero, the bias is smaller (\(|e_B|=0.045\),
\(E(|\hat{e}_B|)=0.047\), expected bias of 0.002). At point C, where
almost the entire distribution is on one side of zero, there is
negligible folding in the distribution of the absolute calibration
error, and therefore negligible overestimation.

\begin{figure}[H]
\centering
\caption{Pointwise bias of the absolute calibration error as a function of the
true error, both standardized by the standard error of the smoother.}
\label{fig:biascurve}
\includegraphics[width=\textwidth]{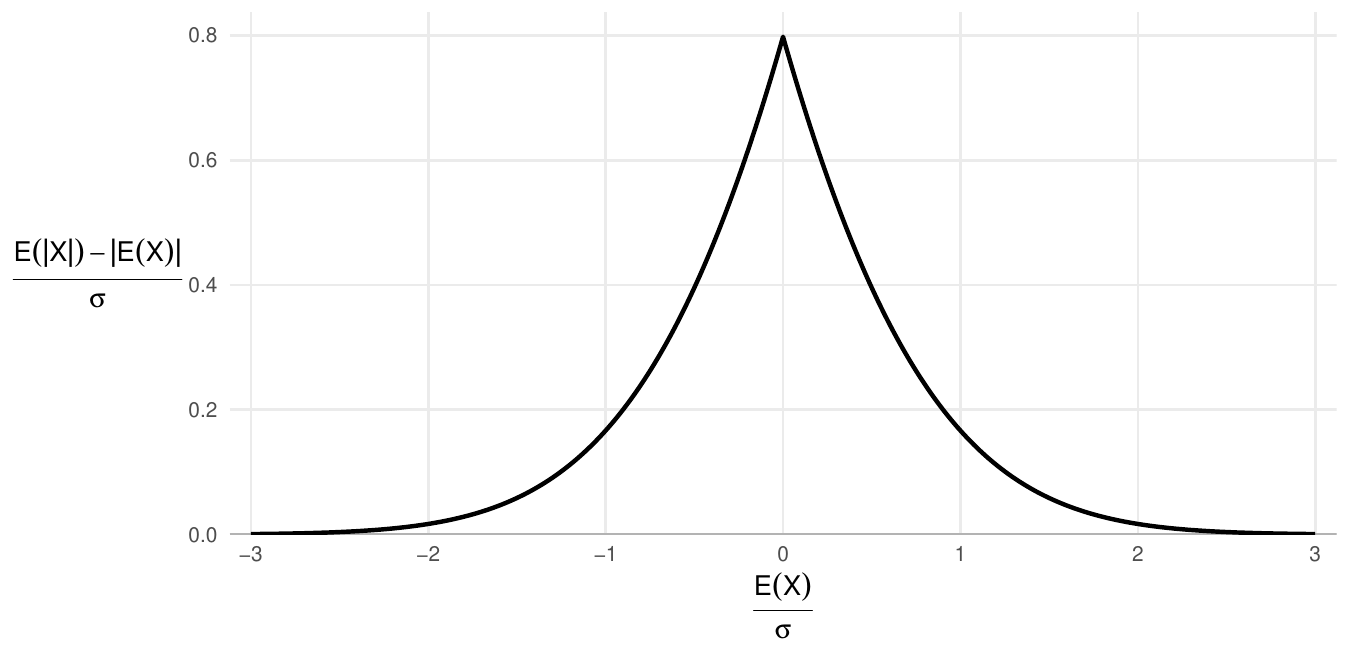}
\par\smallskip
{\footnotesize\raggedright The curve plots $\{\mu_{FN}(e,\sigma)-|e|\}/\sigma$ against $e/\sigma$.\par}
\end{figure}

If the sampling distribution of \(\hat{e}_i\) can be approximated as
normal, which in our context is supported by the Central Limit Theorem,
the magnitude of this bias can be quantified in terms of the
standardized distance (in SE units) of the mean from zero, as shown in
Figure \ref{fig:biascurve}. In this case, \(|\hat{e}_i|\) follows the
folded normal distribution \(FN(e_i,\sigma_i^2)\), and this bias is
\(\mu_{FN}(e_i,\sigma_i)-|e_i|\), where

\[\mu_{FN}(e,\sigma)=\sigma\sqrt{2/\pi}\,\exp\!\left(-\frac{e^{2}}{2\sigma^{2}}\right)+e\left[2\Phi\!\left(\frac{e}{\sigma}\right)-1\right]\]

is the mean of the \(FN(e,\sigma^2)\) distribution (\(\pi\) is the
mathematical constant and \(\Phi(\cdot)\) is the standard normal CDF)
{[}23{]}.

Thus, for a given distance of the calibration curve from the line of
perfect calibration, the bias is larger when the SE is larger. Even if
the true ICI is the same across two groups, the estimated ICI will tend
to be larger in the group with the smaller (effective) sample size.

Note that the folding bias is relative to the regularized (smoothed)
estimator of the conditional risk. The biasedness of the smooth
estimator itself can interact with the folding bias in complex ways. For
example, a downward bias in the underlying smooth estimator might offset
part of the upward bias, or an upward bias might aggravate it. Typical
smoothing methods are approximately unbiased under general conditions
(e.g., the true calibration function being smooth) unless close to the
boundaries of predicted risks.

\section{Bias correction strategies}\label{bias-correction-strategies}

Our goal is to estimate and remove the potential bias in the estimation
of ICI and, by extension, quantile-based metrics. Our proposed
approaches are directly informed by the mechanisms explained above. They
fall into two general categories, which we refer to as \textbf{Plug-in}
and \textbf{Inversion} methods. These approaches apply to any smoothing
method that generates an estimate of the calibration error and SE at
each observation. For methods such as LOESS, RCS, and GAM, approximate
SEs are readily produced in standard implementations, and the Central
Limit Theorem justifies the normality assumption for the sampling
distribution -- the same assumption underlying 95\% confidence bands
around calibration curves. Of note, these approaches do not require any
assumptions on the homoscedasticity of errors or, for ICI, the
correlation structure among errors.

\subsection{Plug-in method}\label{plug-in-method}

This method builds on the frequentist approach to the sampling
distribution: the sample estimates of calibration errors are taken as
their population values, which, combined with the estimated SEs, specify
the distribution of estimated calibration errors for a future replicated
study. We then estimate the expected bias of mean absolute errors at
each point as the average distance between the replicated and current
(population) values. This results in an estimated bias for the \(i\) th
observation as

\[\hat{b}_i = \mu_{FN}(\hat{e}_i,\hat{\sigma}_i)-|\hat{e}_i|.\]

This approach does not guarantee \(\hat{b}_i < |\hat{e}_i|\), so
bias-corrected values \(|\hat{e}_i| - \hat{b}_i\) can be negative, which
occurs whenever \(|\hat{e}_i| < 0.436\,\hat{\sigma}_i\). As such, in the
\textbf{Local Plug-in (LP)} approach, we clip individual bias-corrected
values at zero. This results in the following estimator:

\[\widehat{ICI}_{LP} = \frac{1}{n}\sum_{i=1}^n \max\left(|\hat{e}_i| - \hat{b}_i,\, 0\right). \]

In the \textbf{Global Plug-in (GP)} approach, we do not make such a
correction at the individual level. Instead, we clip the bias-corrected
ICI so that it will not have a negative value:

\[\widehat{ICI}_{GP} = \max\left(\frac{1}{n}\sum_{i=1}^n (|\hat{e}_i| - \hat{b}_i),\, 0\right). \]

The unclipped version of these methods can be considered an extension of
the method proposed by Kumar et al.~for binned ECE {[}16, Appendix G{]}
to flexible calibration curves, with the Monte Carlo approximation
replaced by \(\mu_{FN}\).

\subsection{Inversion approaches}\label{inversion-approaches}

As noted, the plug-in method estimates the pointwise bias by taking the
observed error \(\hat{e}\) as the unknown true error \(e\). As Figure
\ref{fig:biascurve} shows, the bias \(b=\mu_{FN}(e,\sigma)-|e|\) peaks
sharply at \(e\) around zero and falls away on both sides, so wherever
the true error is small relative to its SE (i.e., the point is near the
summit in Figure \ref{fig:biascurve}), replacing \(e\) with \(\hat{e}\)
tends to move down the bias curve, and thus the estimated bias from the
plug-in methods is smaller than the true bias. Of note, the inequality
reverses for larger values of \(|e|\) in the tails of the bias curve,
where the bias function is convex and the estimated bias overestimates
the true one, due to Jensen's inequality. However, in these regions the
bias is already small, so the practical consequence tends to be an
under-correction when the bias is large.

The inversion approach directly addresses this issue by considering the
true calibration errors as latent quantities. It asks the question: what
value of \(|e|\), together with the observed \(\hat \sigma\), generates
the observed \(|\hat{e}|\) by expectation? Because
\(\mu_{FN}(e,\sigma) \ge |e|\), this latent error is never larger than
the observed error, so this method can be formalized as solving for
\(k \in [0,1]\) such that
\(\mu_{FN}(k\hat{e},\hat{\sigma}) = |\hat{e}|\). As the inverse of the
folded normal mean function does not have a closed form, solving this
equation requires univariate root-finding. Of note, no solution exists
when \(|\hat{e}|\) is lower than the noise floor
(\(\mu_{FN}(0, \hat{\sigma}_i) = \hat{\sigma}_i\sqrt{2/\pi}\)), in which
case we set \(k=0\) as the value that minimizes the distance to the
observed error.

In the \textbf{Local Inversion (LI)} approach, such inversion is applied
pointwise, and the corrected ICI is the average of the latent absolute
errors \(k_i|\hat{e}_i|\):
\[\widehat{ICI}_{LI} = \frac{1}{n}\sum_{i=1}^n k_i|\hat{e}_i|, \] where
\(k_i\) is the unique value between 0 and 1 such that
\(\mu_{FN}(k_i\hat{e}_i,\hat{\sigma}_i) = |\hat{e}_i|\) or \(0\) when
\(|\hat{e}_i|\le\mu_{FN}(0,\hat{\sigma}_i) = \hat{\sigma}_i\sqrt{2/\pi}\).

Instead of estimating this inverse for all points (i.e., solving for
\(n\) parameters), one can impose constraints. One approach is to
consider the multiplier factors to be equal across all points,
\(k_i = k\), i.e., considering the true calibration error function to be
a scaled version of the observed one. By solving for a unique scaling
value we effectively borrow strength across points, which might result
in more efficient bias correction (as well as faster numerical
implementation). We call this approach \textbf{Global Inversion (GI)},
as it applies one correction factor across all points:
\[\widehat{ICI}_{GI} = \frac{1}{n}\sum_{i=1}^n k|\hat{e}_i|, \] where
\(k\) is the unique value such that
\(\sum_{i=1}^n \mu_{FN}(k|\hat{e}_i|,\hat{\sigma}_i) = \sum_{i=1}^n |\hat{e}_i|\),
with \(k=0\) when \(\widehat{ICI}\) does not exceed the noise floor
\(\frac{1}{n}\sum_{i=1}^n \hat{\sigma}_i\sqrt{2/\pi}\). This requires
only one univariate root-finding to find the value of \(k\).

\section{An illustrative example}\label{sec-example}

We use the GUSTO-I trial data {[}24{]}, a dataset commonly used for
methodological research in risk prediction modeling {[}25,26{]}. The
primary endpoint is 30-day mortality in patients with acute myocardial
infarction. The candidate prediction model is similar to those used
previously {[}8{]}, and contains the following seven variables: age,
shock (Killip class III or IV), anterior infarct location or previous
myocardial infarction, diabetes, hypotension (systolic blood pressure
below 100 mmHg), tachycardia (heart rate above 80/min), and sex. The
total sample size of GUSTO was 40,830, with 2,851 events (7.0\%).

We randomly divide this dataset into a development sample of size 8,830,
and leave the remaining 32,000 as the total validation sample to draw
the flexible calibration curve and compute (bias-corrected) ICIs. Two
risk prediction models are fitted using unpenalized logistic regression:
one on the entire development sample, which is expected to be
well-calibrated in the validation sample due to the random splitting of
the data and the large development sample, and one on a random 500 of
its patients, which is expected to be miscalibrated due to the small
development sample. Each model is assessed in validation samples of
sizes 500 and 32,000.

\begin{figure}[H]
\centering
\caption{Estimated calibration curves (top) and absolute calibration error with different bias correction methods (bottom). ICI values are multiplied by $10^3$.}
\label{fig:gustoex}
\includegraphics[width=\textwidth]{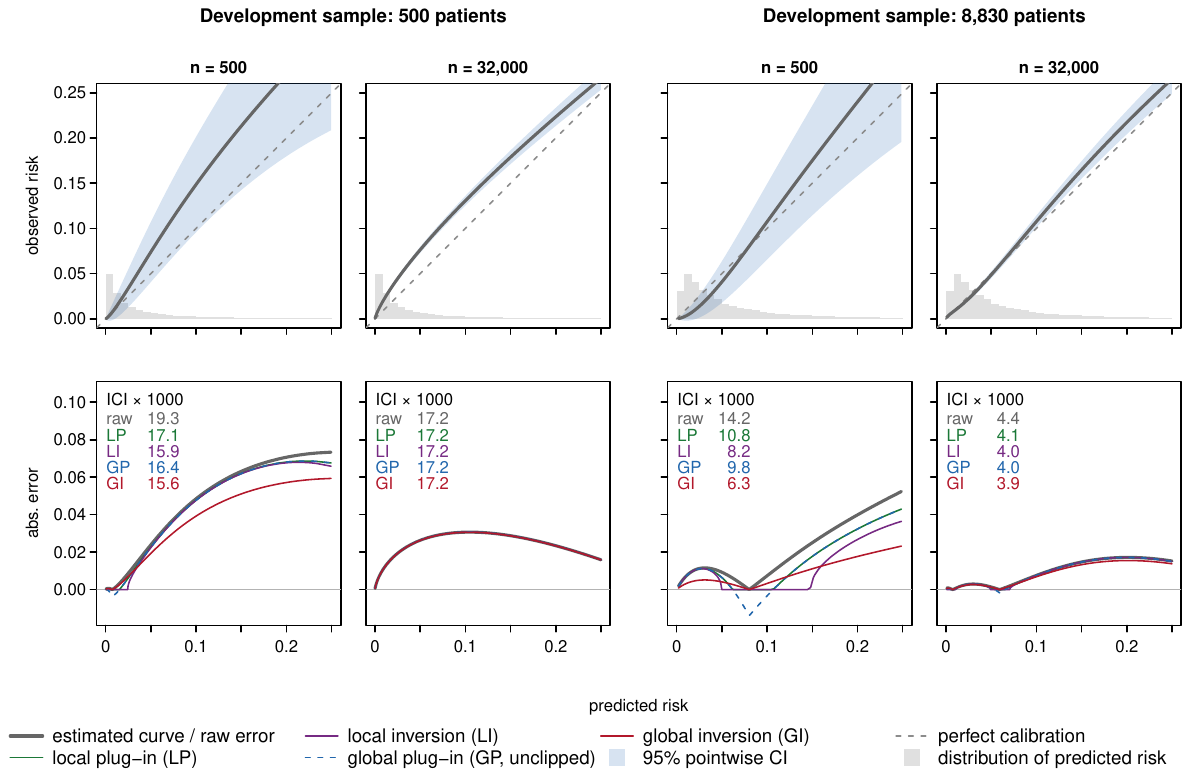}
\par\smallskip
{\footnotesize\raggedright Calibration curves are estimated by a logistic generalized additive model; shaded bands are 95\% pointwise confidence intervals ($\pm 1.96$ SE); dashed line: perfect calibration; grey bars: distribution of predicted risks. 

ICI: integrated calibration index; LP: local plug-in; GP: global plug-in; LI: local inversion; GI: global inversion.\par}
\end{figure}

Figure \ref{fig:gustoex} shows the estimated calibration curves (based
on a GAM smoother) for each model at each validation sample size. The
bottom panels depict the observed absolute calibration error (grey
curves) and their bias-corrected versions using the four methods.

The bias-corrected error curves illustrate how each method operates. For
the plug-in methods, these curves are based on subtracting the expected
bias from the observed error curve. For the inversion approaches, they
are the assumed true error curves from which the observed curve arises
by expectation. When the curve is far away from the line of perfect
calibration (as in the second column), the raw and bias-corrected ICIs
are the same. On the other hand, when the curve is close to the line of
perfect calibration in SE units (as in the third column), the correction
is most pronounced. Note that the LP and GP methods only differ in
whether they tolerate negative absolute error estimates, with LP
clipping negative values to 0 at each point. The LI curve is zero at any
point where the observed absolute error is below
\(\mu_{FN}(0,\hat{\sigma}_i)\). The GI curve is always a scaled version
of the observed curve.

\section{Extension to other metrics}\label{extension-to-other-metrics}

The concepts explained above can in principle be applied to other
calibration metrics, including quantile-based ones (e.g., E50 and E90)
that are also used in clinical prediction modeling. A direct extension
of these methods for such metrics is conceivable. For example, for E50
and the plug-in approach, one would construct a sampling distribution
for replicated calibration curves centered on the observed curve and
compute the expected value of the median absolute errors from future
replicated curves. However, unlike ICI, which is a sum-type metric whose
expectation depends only on the marginal distribution of pointwise
calibration errors, the expected value of a quantile depends on the
joint distribution of the errors along the curve. One should either
incorporate such dependence structure explicitly (e.g., via the reported
covariance matrix of the estimated calibration curve) or resort to
methods such as bootstrapping. These are operationally feasible, but are
against the desire for simplicity in our proposal (and bootstrapping
introduces Monte Carlo noise).

Instead, one can work with the mean-matched error curves that are
constructed for the ICI. In this approach, quantile-based metrics are
directly derived from the mean-matched absolute error curves (see bottom
panels of Figure \ref{fig:gustoex} for such curves). Any summary
statistic of calibration error, including E50 and E90, is derivable from
these curves, so that one bias correction strategy generates all summary
statistics of interest. Because the curves match the mean rather than a
quantile, the bias of the derived summaries is not targeted directly. In
the simulation studies described below, we investigate how mean-matched
bias correction performs for quantile-based metrics.

\section{Simulation studies}\label{simulation-studies}

We performed simulation studies to evaluate the performance of these
approaches in reducing the bias of estimated ICI and quantile metrics
under various degrees of miscalibration. We adapted the general
simulation design by Austin and Steyerberg in their original publication
introducing the ICI {[}9{]}, with modifications that enable computing
the true ICI and introducing scenarios where the model is truly
calibrated. As in Austin and Steyerberg, two simulations were conducted,
where the true model contained, respectively, a quadratic and an
interaction term.

For the `quadratic' simulation setting, the response-generating model
was of the form \(\text{logit}(P(Y=1))=x + a_2x^2\), with \(x\) having a
standard normal distribution. The prediction model was a logistic model
involving \(x\) only, omitting the quadratic term. In each replicate,
this model was fitted to an independent development sample of 1,000
individuals and then evaluated in a random validation sample of size
\(n\), in which ICI was estimated. Unlike in Austin and Steyerberg
(where development and validation samples grow together), the
development sample size was held fixed so that the population of
prediction models, and hence of true ICIs, was the same at every
validation sample size. Using a factorial design, we varied the sample
size, \(n \in \{250, 500, 1000, 2000, 4000\}\), and the strength of the
omitted term, \(a_2 \in \{0, 0.25, 0.5\}\). Note that \(a_2=0\) results
in the prediction model being correctly specified, although still
subject to estimation error, and thus always with some degree of
miscalibration. As such, in addition to these values of \(a_2\), we
considered a new scenario in which the prediction model was the true
model itself (thus true ICI = 0).

In the `interaction' simulation setup, the true response-generating
model was \(\text{logit}(P(Y=1))=x_1 + x_2 + a_3 x_1 x_2\), with \(x_1\)
and \(x_2\) having independent standard normal distributions. Here, the
prediction model omitted the interaction term, and was again fitted in
an independent development sample of 1,000 individuals in each
replicate, with an extra scenario where the prediction model was set to
be the same as the true model. Similar to the previous simulations, for
any value of \(n\), a validation sample of size \(n\) was obtained to
estimate and compare ICIs. The sample size,
\(n \in \{250, 500, 1000, 2000, 4000\}\), and the strength of the
omitted term, \(a_3 \in \{0, 0.5, 1\}\), were varied in a factorial
design.

The values of \(a_2\) and \(a_3\) were chosen so that the two designs
span plausible and comparable degrees of miscalibration (mean true ICI
of the miscalibrated models from 15 to 80 \(\times 10^{-3}\) in the
quadratic and from 14 to 57 \(\times 10^{-3}\) in the interaction
design). The true calibration curves of the prediction models are shown
in Appendix \ref{app-devcurves}.

Each scenario was replicated 1,000 times. Raw ICI, true ICI, as well as
bias-corrected ICI using each of the four methods were computed. The
true ICI of each fitted model was computed exactly, as the average over
the covariate distribution of \(|g(\hat{P}) - \hat{P}|\), where
\(g(\hat{P})\) is the true risk among individuals with predicted risk
\(\hat{P}\) (by numerical integration). We investigated the performance
of these methods in terms of average bias across simulations
(\(E[\widehat{ICI} - ICI]\)), as well as mean absolute error (MAE,
\(E|\widehat{ICI} - ICI|\)). MAE was chosen over mean squared error as
it is more aligned with ICI being an L1 quantity. The figures in the
main text pertain to ICI with the LOESS smoother; all other results are
provided in Appendices \ref{app-smoothers} and \ref{app-quantiles}.

\begin{figure}[H]
\centering
\caption{Bias and mean absolute error of the raw and bias-corrected ICIs against
validation sample size, quadratic design (all values are multiplied by 1000).}
\label{fig:simquad}
\includegraphics[width=\textwidth]{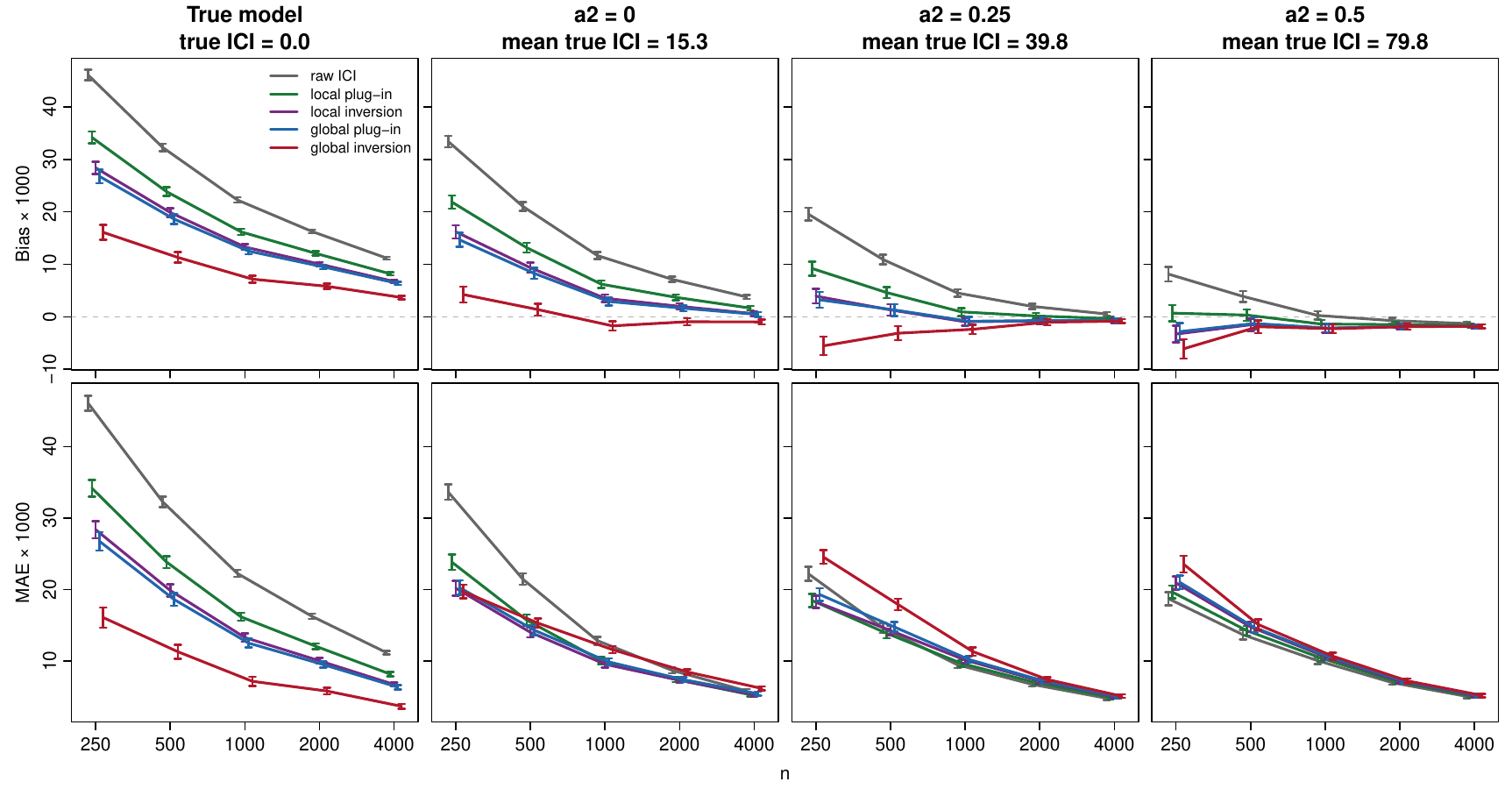}
\par\smallskip
{\footnotesize\raggedright Top row: bias; bottom row: mean
absolute error (MAE). Vertical bars are 95\% Monte Carlo
intervals; points are perturbed horizontally for better legibility. Calibration curves are estimated by LOESS (span
0.75, local quadratic). ICI: integrated calibration index; LOESS: locally estimated scatterplot smoothing; LP: local plug-in; GP: global plug-in; LI: local inversion; GI: global inversion.\par}
\end{figure}

Results for the quadratic simulations are provided in Figure
\ref{fig:simquad}. The first observation is a large upward bias in raw
ICI, especially with small sample sizes and models that are calibrated
or moderately miscalibrated -- a pattern that, while attenuated with
sample size, was still noticeable even at \(n=4000\). Raw ICI decreased
by more than half where the model was calibrated or mildly miscalibrated
(the first two columns) with increasing sample size. The results broadly
agree with the theoretical expectation that bias correction matters the
most where true ICI is close to zero. Relative reduction in bias was
substantial even at \(n=4000\) in these scenarios. All four methods
decreased the bias of raw ICI across all scenarios and sample sizes.
When the model was calibrated or slightly miscalibrated, the GI method
substantially improved bias and MAE. When the miscalibration was
moderate (\(a_2=0.25\)), bias-corrected ICIs performed better than the
raw ICI in terms of bias, even though GI overcorrected the bias,
resulting in it having slightly higher MAE.

\begin{figure}[H]
\centering
\caption{Bias and mean absolute error of the raw and bias-corrected ICIs against
validation sample size, interaction design (all values are multiplied by 1000).}
\label{fig:siminter}
\includegraphics[width=\textwidth]{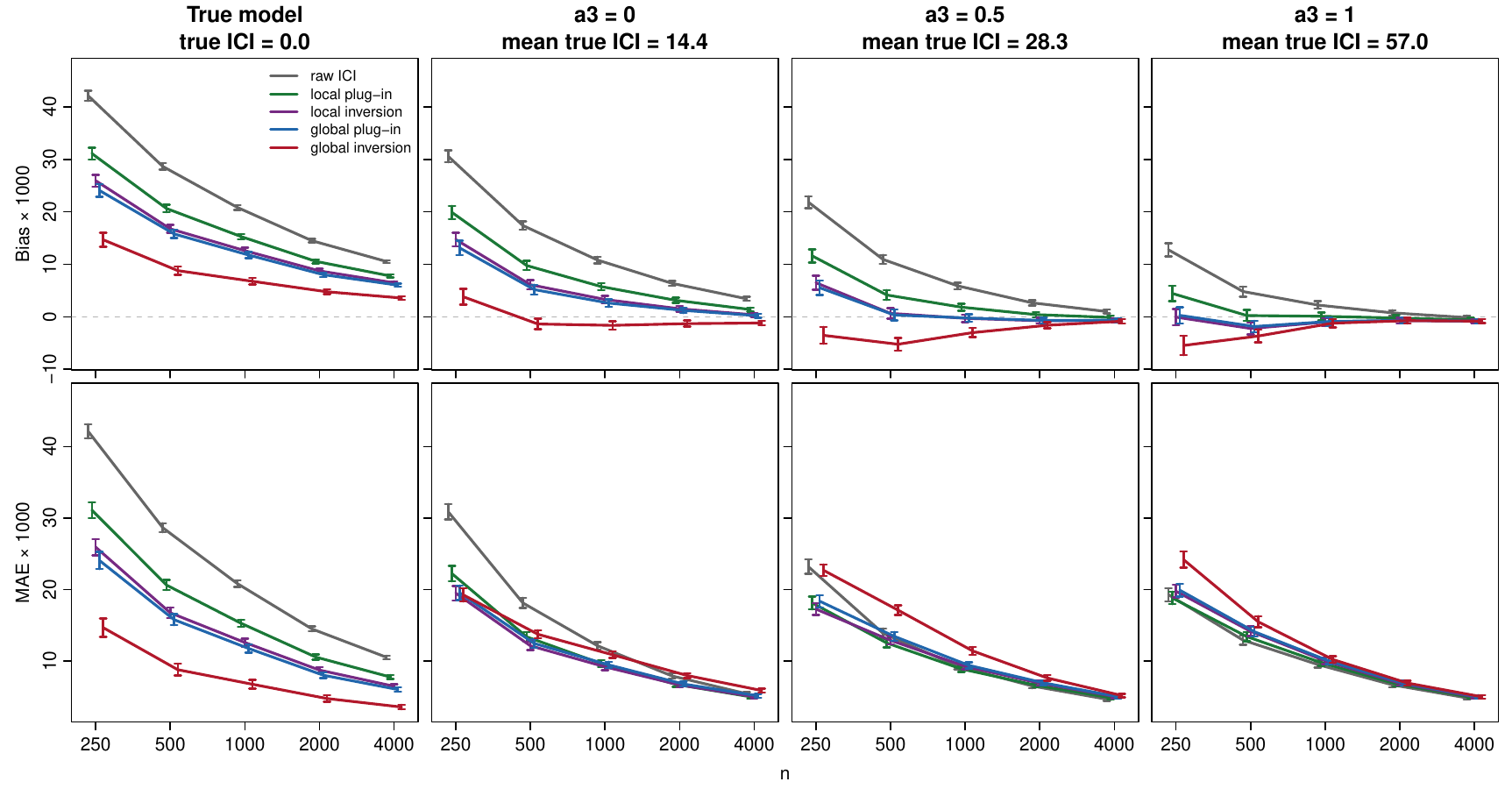}
\par\smallskip
{\footnotesize\raggedright Same layout, replicates, smoother and
conventions as Figure \ref{fig:simquad}; ICI: integrated calibration index; MAE: mean absolute error; LOESS:
locally estimated scatterplot smoothing; LP: local plug-in; GP: global plug-in; LI: local inversion; GI: global inversion.\par}
\end{figure}

Results for the interaction simulation are provided in Figure
\ref{fig:siminter} and they broadly resemble those from the first set of
simulations. Here, with perfect calibration or mild miscalibration,
where the upward bias is the most substantial, GI remained the best in
terms of bias and MAE. Again, bias correction methods reduced bias
compared with raw ICI, with the GI method overcorrecting slightly in the
moderate and large miscalibration scenarios (right-most two columns).

Details for other smoothers (RCS and GAM) are provided in Appendix
\ref{app-smoothers}. The overall conclusions for ICI are broadly similar
to those for LOESS.

Results for E50 and E90 for all three smoothers are provided in Appendix
\ref{app-quantiles}. Again, overall, the GI method outperformed others
substantially when the upward bias was large, but it could have a
slightly higher MAE when the bias was small. For E50, a single bias
correction method fine-tuned for ICI could largely mitigate the bias,
with findings on the performance of each approach being broadly similar
to their corresponding performance for ICI. For E90, however, the GI
method was substantially better than other approaches in correcting the
bias of the raw estimates. The local methods (and GP, whose values equal
those of LP) generated estimates that were generally indistinguishable
from the raw E90.

To summarize the output of simulations, we aggregated the scenario-level
results in Table \ref{tab:summary}, separately for the true model and
for the miscalibrated models. For each method and metric (ICI, E50, and
E90), we computed the average absolute bias, and reported the relative
reduction compared with the raw method. The same was done for MAE.

\begin{table}[htbp]
\centering
\caption{Bias and mean absolute error (MAE) of the raw and bias-corrected ICI, E50 and E90, averaged over simulation scenarios (values are multiplied by 1000).}
\label{tab:summary}
\small
\begin{tabular}{@{}llrrrrr@{}}
\hline
Summary & Smoother & Raw & LP & LI & GP & GI \\
\hline
\multicolumn{7}{@{}l}{\emph{A. True model (true value of 0): mean estimate, equal to bias and MAE}} \\
ICI & LOESS & 24.5 & 18.0 ($-$27\%) & 14.9 ($-$39\%) & 13.9 ($-$43\%) & 8.3 ($-$66\%) \\
 & RCS & 23.0 & 17.4 ($-$24\%) & 14.6 ($-$36\%) & 14.2 ($-$38\%) & 9.5 ($-$59\%) \\
 & GAM & 18.4 & 14.0 ($-$24\%) & 11.8 ($-$36\%) & 12.1 ($-$34\%) & 8.9 ($-$52\%) \\
\addlinespace[1pt]
E50 & LOESS & 22.2 & 13.7 ($-$38\%) & 8.3 ($-$63\%) & = LP & 7.8 ($-$65\%) \\
 & RCS & 21.5 & 14.4 ($-$33\%) & 9.6 ($-$56\%) & = LP & 9.2 ($-$57\%) \\
 & GAM & 18.8 & 13.9 ($-$26\%) & 10.7 ($-$43\%) & = LP & 9.2 ($-$51\%) \\
\addlinespace[1pt]
E90 & LOESS & 46.2 & 43.1 ($-$7\%) & 41.0 ($-$11\%) & = LP & 15.0 ($-$68\%) \\
 & RCS & 42.9 & 40.0 ($-$7\%) & 38.1 ($-$11\%) & = LP & 17.0 ($-$60\%) \\
 & GAM & 31.0 & 27.9 ($-$10\%) & 25.9 ($-$16\%) & = LP & 14.5 ($-$53\%) \\
\addlinespace[4pt]
\multicolumn{7}{@{}l}{\emph{B. Miscalibrated models: mean absolute bias}} \\
ICI & LOESS & 8.7 & 4.4 ($-$50\%) & 3.0 ($-$65\%) & 2.6 ($-$70\%) & 2.4 ($-$72\%) \\
 & RCS & 10.7 & 7.0 ($-$35\%) & 5.1 ($-$52\%) & 5.0 ($-$53\%) & 2.9 ($-$73\%) \\
 & GAM & 5.2 & 2.7 ($-$48\%) & 2.8 ($-$46\%) & 2.9 ($-$44\%) & 3.0 ($-$42\%) \\
\addlinespace[1pt]
E50 & LOESS & 6.5 & 2.6 ($-$60\%) & 2.9 ($-$56\%) & = LP & 3.5 ($-$46\%) \\
 & RCS & 9.7 & 6.1 ($-$37\%) & 4.2 ($-$57\%) & = LP & 3.2 ($-$67\%) \\
 & GAM & 4.6 & 2.7 ($-$42\%) & 2.6 ($-$43\%) & = LP & 3.1 ($-$32\%) \\
\addlinespace[1pt]
E90 & LOESS & 19.2 & 17.7 ($-$8\%) & 16.9 ($-$12\%) & = LP & 3.1 ($-$84\%) \\
 & RCS & 20.7 & 19.7 ($-$5\%) & 19.1 ($-$8\%) & = LP & 5.7 ($-$73\%) \\
 & GAM & 9.9 & 8.7 ($-$12\%) & 8.0 ($-$19\%) & = LP & 5.1 ($-$49\%) \\
\addlinespace[4pt]
\multicolumn{7}{@{}l}{\emph{C. Miscalibrated models: mean absolute error}} \\
ICI & LOESS & 12.7 & 11.2 ($-$12\%) & 11.0 ($-$13\%) & 11.2 ($-$12\%) & 12.4 ($-$2\%) \\
 & RCS & 13.2 & 11.7 ($-$11\%) & 11.4 ($-$14\%) & 11.5 ($-$13\%) & 12.3 ($-$7\%) \\
 & GAM & 10.9 & 10.7 ($-$2\%) & 11.0 (0\%) & 11.0 (+1\%) & 12.0 (+10\%) \\
\addlinespace[1pt]
E50 & LOESS & 13.3 & 13.4 (+1\%) & 15.0 (+13\%) & = LP & 13.6 (+3\%) \\
 & RCS & 13.8 & 13.7 ($-$1\%) & 14.7 (+6\%) & = LP & 13.3 ($-$4\%) \\
 & GAM & 12.3 & 12.9 (+5\%) & 14.1 (+15\%) & = LP & 13.2 (+7\%) \\
\addlinespace[1pt]
E90 & LOESS & 24.7 & 24.2 ($-$2\%) & 24.3 ($-$1\%) & = LP & 21.8 ($-$12\%) \\
 & RCS & 24.5 & 24.2 ($-$1\%) & 24.3 ($-$1\%) & = LP & 21.2 ($-$13\%) \\
 & GAM & 18.7 & 18.9 (+1\%) & 19.3 (+3\%) & = LP & 19.9 (+6\%) \\
\hline
\end{tabular}

\par\smallskip
{\footnotesize\raggedright Values ($\times 1000$) are averaged over
simulation scenarios, with the percent change from the raw estimate in parentheses
(negative: better). Panel A: the 10 scenarios in which the prediction model is the
correct model (true ICI, E50 and E90 of zero); as true ICI is 0, the average absolute bias is equal to MAE. Panels B and
C: the 30 scenarios with miscalibrated models. For bias, absolute values are averaged so that over- and
under-estimation do not cancel out. For E50 and E90, GP and LP estimates are identical, as clipping at zero before or after taking a quantile gives the same quantile-based result. ICI: integrated calibration index; E50, E90: median and 90th percentile of the
absolute calibration error; MAE: mean absolute error; LOESS: locally estimated
scatterplot smoothing; RCS: restricted cubic spline; GAM: generalized additive
model; LP: local plug-in; LI: local inversion; GP: global plug-in; GI: global
inversion.\par}
\end{table}

\section{Empirical evaluation in GUSTO
data}\label{empirical-evaluation-in-gusto-data}

Here, we use the GUSTO data to further explore the performance of these
methods. While a real-world dataset does not offer an opportunity to
compare the estimates against the truth, the large sample enables the
comparison of ICI estimated in widely different samples. With a fixed
fitted model, changes in estimates across validation sample sizes
indicate the extent of the bias.

The design broadly follows the illustrative example in Section
\ref{sec-example}. Within each replicate, we split the data at random
into a validation pool of 32,000 patients and a development set of the
remaining 8,830. Two prediction models, each a logistic regression on
the same seven main effects, were fitted on that development data and
then frozen: one using a random 500 of the development patients, with an
events-per-variable (EPV) ratio of \textasciitilde5 (referred to as
\emph{Small Development Sample}), and one using all 8,830, corresponding
to an EPV of \textasciitilde88 (referred to as \emph{Large Development
Sample}). These models are assessed in random validation subsamples of
increasing size, drawn independently from the validation pool
(\(n \in \{500, 2000, 8000, 32000\}\)). The procedure was repeated 100
times.

Because the model and the validation pool are held fixed across varying
sample sizes, the estimand does not change with the validation sample
size. As such, any systematic movement with \(n\) is a manifestation of
estimation bias. An algorithm performs well here if its expected value
drifts little with the sample size and converges to the same value as
\(n\) grows.

Table \ref{tab:gusto} reports the results. In the large-development
scenario, the raw ICI falls from 13.0 to 4.4 per thousand as the
validation sample grows, even though the estimand is identical
throughout; at \(n = 500\) it is barely distinguishable from the noise
floor a perfectly calibrated model would produce at that sample size
(\(12.0 \times 10^{-3}\)), so an analyst would report substantial
miscalibration for a model that is actually well calibrated. All methods
reduce that bias. In the small-development scenario, where the model is
genuinely poorly calibrated (given the high ICIs with large validation
samples), the signal sits well above the floor from \(n = 2000\) onwards
and correction matters much less there; but even here, at \(n=500\) the
correction is still noticeable. Again, these results indicate that bias
correction is most consequential where true bias is large compared with
true ICI -- the situations where raw ICI is at risk of exaggerating the
miscalibration.

\begin{table}[htbp]
\centering
\caption{Raw and bias-corrected ICIs in the GUSTO case study, by development
scenario and validation sample size.}
\label{tab:gusto}
\begin{tabular}{@{}lrrrrr@{}}
\hline
$n$ & Raw ICI & LP & LI & GP & GI \\
\hline
\multicolumn{6}{@{}l}{\emph{Small Development Sample (500 patients)}} \\
500 & 19.4 (0.9) & 17.1 (0.9) & 15.8 (1.0) & 16.0 (1.0) & 14.4 (1.1) \\
2,000 & 17.0 (0.9) & 16.2 (0.9) & 15.8 (0.9) & 15.8 (0.9) & 15.5 (1.0) \\
8,000 & 16.4 (0.8) & 16.2 (0.9) & 16.1 (0.9) & 16.1 (0.9) & 16.0 (0.9) \\
32,000 & 15.9 (0.8) & 15.8 (0.8) & 15.8 (0.8) & 15.8 (0.8) & 15.8 (0.8) \\
\addlinespace
\multicolumn{6}{@{}l}{\emph{Large Development Sample (8,830 patients)}} \\
500 & 13.0 (0.8) & 10.0 (0.8) & 8.6 (0.8) & 8.6 (0.8) & 6.7 (0.9) \\
2,000 & 7.1 (0.4) & 5.7 (0.4) & 5.0 (0.4) & 5.1 (0.4) & 4.1 (0.4) \\
8,000 & 4.8 (0.2) & 4.2 (0.3) & 3.8 (0.3) & 3.8 (0.3) & 3.3 (0.3) \\
32,000 & 4.4 (0.1) & 4.1 (0.2) & 4.0 (0.2) & 4.0 (0.2) & 3.9 (0.2) \\
\hline
\end{tabular}

\par\smallskip
{\footnotesize\raggedright \textit{Note:} A logistic model with seven main
effects is fitted once on a development sample, then assessed in random
subsamples of a fixed validation pool of 32,000 patients at
4 validation sample sizes $n$. Entries are the mean over 100 replicates in
units of $10^{-3}$, with the Monte Carlo standard error of that mean in parentheses.
GUSTO: Global Utilization of Streptokinase and Tissue Plasminogen Activator for
Occluded Coronary Arteries; ICI: integrated calibration index; LP: local plug-in; LI:
local inversion; GP: global plug-in; GI: global inversion.\par}
\end{table}

\section{Discussion}\label{discussion}

The contributions from this paper can be summarized as follows: first,
we conceptualized the upward bias in the summary metrics of absolute
calibration error as the `folding' phenomenon: the bias due to taking
the absolute value of a noisy estimate whose distribution spans zero. We
showed that this bias can be substantial, with the degree of bias
generally depending on the distance, in SE units, of the calibration
curve from the line of perfect calibration. Second, we studied two
general approaches for mitigating this bias. While the plug-in methods
can be seen as an extension of previous methods for binned calibration
{[}16{]}, the inversion methods are, to the best of our knowledge, new
for summary metrics of miscalibration of prediction models. Our
numerical results indicate that the proposed methods generally mitigate
the upward bias, with little or no cost in the mean absolute error. In
our numerical experiments, the \emph{Global Inversion} method generally
outperformed the others, particularly where bias was substantial and
bias correction mattered the most. The same mean-matched curves also
reduced the bias of quantile-based metrics (albeit, for E90, only Global
Inversion performed substantially better than raw values). An attractive
feature of these approaches is that they only need an estimate of the
calibration curve and its sampling variance at each observation. As
such, they are agnostic to the method used to construct the calibration
curve for different outcomes (e.g., binary, ordinal, survival, and
counterfactual). This also makes these methods generally easy to
implement, with the Global Inversion method requiring a single
univariate root-finding of a smooth, monotone, deterministic function. A
basic R implementation of this approach is provided in Appendix
\ref{app-code}.

These findings are particularly important when comparing ICI values of a
model across subgroups of different sizes, for example as a fairness
metric. Indeed, calibration is considered a fairness criterion in the
algorithmic fairness literature {[}19--21{]}, and ICI-type metrics have
been used as objective summaries for comparing calibration across
subgroups {[}27,28{]}. Generally, if the actual ICIs are similar among
subgroups, the bias in ICI and quantile-based metrics tends to be larger
in the smaller subgroup. Marginalized groups are often under-represented
in both development and validation data. A sample-size-dependent upward
bias in a metric like ICI can obscure the legitimate concerns that
predictions might be miscalibrated among under-represented groups due to
their smaller share in the development sample. In one such comparison,
the smallest racial and ethnic group was more than ten times smaller
than the largest {[}27{]}; at such imbalances, the smaller group's
estimated calibration error can be substantially inflated by folding
alone. Raw ICI gaps thus conflate differential calibration with
differential sample size. Bias correction approaches such as GI would
make such subgroup comparisons more meaningful.

Much of the previous literature on mitigating the bias of miscalibration
summaries, mainly in meteorology and machine learning, has focused on
summaries based on quadratic errors {[}14--16,29--31{]}. For quadratic
measures of miscalibration (such as average squared calibration error),
the bias is directly related to the variance of the smooth estimators
(as \(E(\hat{e}^2)=e^2+\sigma^2\)) and thus can easily be quantified
{[}15,16{]}. For measures of absolute errors, Roelofs et al.~showed that
the bias of binned ECE also depends on the binning scheme {[}17{]}.
Kernel-based alternatives to binning have been proposed, partly to
reduce this bias {[}32,33{]}, and Kumar et al. {[}16{]} suggested a
Monte-Carlo-based correction for this bias, which generally corresponds
to the (unclipped) plug-in approach proposed here. Recently, Berta et
al.~proposed a variational approach for estimating ECE based on
cross-validation (repeated fitting of the recalibration model in the
training subsample to predict the direction of prediction error in the
held-out subsample) {[}18{]}. This approach does not satisfy our
definition of a general bias mitigation method as it does not extend to
quantile-based metrics and to partially observed outcome types such as
survival or counterfactual ones. Further, unless leave-one-out
cross-validation is used, this approach might be sensitive to the number
of folds. On the other hand, as it only requires the sign of the
calibration error from the smoother, it might be more robust to the bias
of the smoother itself compared with our approaches.

There are several areas of future research. First, the local and global
inversion methods can be seen as two ends of a spectrum: one can
consider an inversion multiplier \(k\) that changes across the range of
the calibration curve but has similar values in nearby regions. For
example, a binned version would estimate separate \(k\)s within \(d\)
bins. With \(d=1\), this reduces to the global inversion method, and
with \(d=n\) it will be the local inversion method. We did not study
this method as it introduces a tuning parameter whose default value
needs to be assessed separately. As mentioned earlier, the proposed
methodology can be generalized to any outcome type, including ordinal
values {[}34{]}, right-censored outcomes {[}35{]}, and counterfactual
predictions {[}4,13{]}. However, the bias-reduction performance of these
methods for such outcomes should be investigated in dedicated studies.
As expected, simulations consistently showed that these methods perform
best when the calibration curve is close to the identity line in SE
units; however, far from the identity line, these approaches can
slightly increase the mean absolute error. This makes it possible for
future research to produce decision rules as to when a correction needs
to be invoked.

An interesting implication of the proposed approaches concerns
uncertainty quantification around ICI and quantile-based metrics. These
metrics are often reported with a 95\% confidence interval (CI),
conventionally obtained by the bootstrap. First, bootstrap replicates of
ICI are themselves subject to folding, which shifts such intervals
upwards. A bias-corrected confidence interval can be obtained by
incorporating one of the proposed bias correction methods inside the
bootstrap. Further, a conceptual problem with the general
bootstrap-based approach is that the null value of these metrics (0) is
never inside the interval (even if the model is calibrated), because
\(\widehat{ICI}\) is an average of absolute errors and is positive with
probability 1 (and similarly for quantile-based metrics). As GP and GI
can return exactly zero ICIs, bias-corrected ICIs might be zero in a
fraction of bootstrap replicates, with this fraction increasing as the
true ICI approaches zero. If at least a fraction \(\alpha/2\) of
bootstrap replicates is zero (e.g., 2.5\% for a 95\% CI), the
\(100(1-\alpha)\%\) percentile CI includes the null value and can be
considered compatible with the model being moderately calibrated. The
coverage of such bias-corrected, null-compatible CIs should be
investigated in dedicated studies.

In summary, the upward bias in ICI and similar summaries of absolute
calibration error can be substantial, and can differentially affect
calibration assessment in subgroups of different sizes. This is
particularly concerning in fairness assessments of predictive
algorithms, as raw summaries of calibration can result in an unfair
assessment of a model's fairness. The proposed approaches reduce the
bias substantially when it is large, at the cost of at most a modest
increase in mean absolute error when it is small. We suggest that
whenever investigators report a flexible calibration curve, the
accompanying summary statistics be bias-adjusted using an approach that
builds directly on the reported curve and its associated uncertainty.
Global Inversion is our recommended method, given its ease of
implementation and performance in reducing the bias of ICI and
quantile-based metrics.

\clearpage

\section*{References}\label{references}
\addcontentsline{toc}{section}{References}

\protect\phantomsection\label{refs}

\clearpage

\appendix

\section{Calibration of the simulated prediction
models}\label{app-devcurves}

\begin{figure}[H]
\centering
\caption{True calibration curves of the prediction models in the simulation
studies.}
\label{fig:devcurves}
\includegraphics[width=\textwidth]{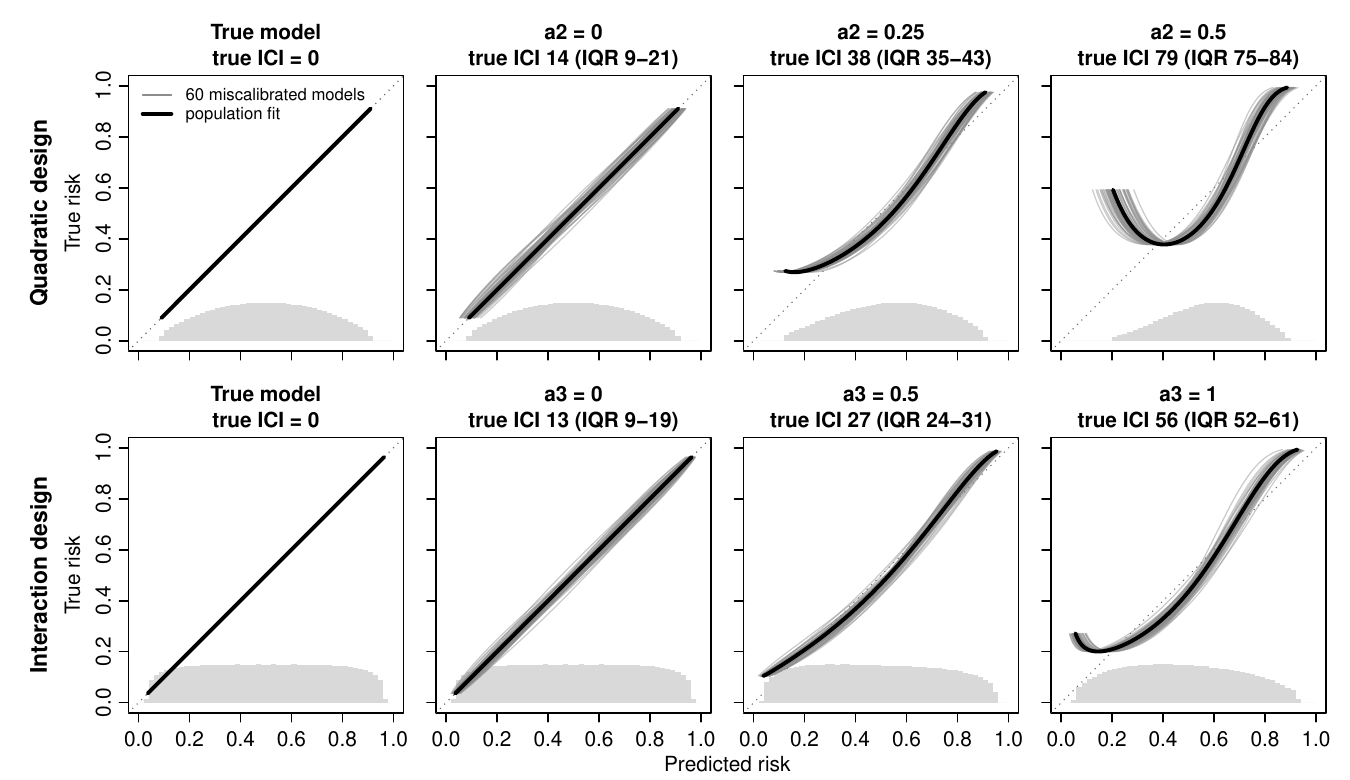}
\par\smallskip
{\footnotesize\raggedright \textit{Note:} Grey: 60 randomly chosen miscalibrated models,
each developed on 1,000 individuals; black: the model fitted to an infinite
development sample. Dotted line: perfect calibration. ICI: integrated calibration
index. Each curve is drawn over the 1st to 99th percentile of its own predicted risks; the shading at the bottom of
each panel shows the distribution of predicted risk. Column headings give the
median and interquartile range of the true ICI ($\times 1000$) across the
1,000 miscalibrated models.\par}
\end{figure}

Figure \ref{fig:devcurves} shows the true calibration curves of the
prediction models in the two simulation designs (top: quadratic design,
bottom: interaction design). The left column is the true model, and so
is perfectly calibrated. Each other column shows 60 randomly chosen
miscalibrated models for each scenario, each developed on 1,000
individuals (grey), together with the model an infinite development
sample would give (black, i.e.~population calibration curve). The dotted
line is the line of perfect calibration.

\clearpage

\section{Simulation results by smoother}\label{app-smoothers}

This section presents the simulation results for two other smoothers,
illustrated in the same layout as that of Figures \ref{fig:simquad} and
\ref{fig:siminter}: logistic regression on a restricted cubic spline in
the logit of the predicted probability (four knots at Harrell's default
quantiles), and a logistic generalized additive model (penalized spline
of the logit of predicted risk, smoothing parameter by restricted
maximum likelihood).

Each point summarizes 1,000 replicates, and all values are multiplied by
1000. All estimators are computed on the same replicates. Across the
three smoothers, the Monte Carlo standard error of a difference from the
raw ICI is at most 0.35 for bias and 0.68 for MAE, on the same scale.
For the true model the true ICI is zero and no estimate is negative, so
bias and MAE have the same value.

\begin{figure}[H]
\centering
\caption{Simulation results, restricted cubic spline, quadratic design.}
\label{fig:rcsquad}
\includegraphics[width=0.9\textwidth]{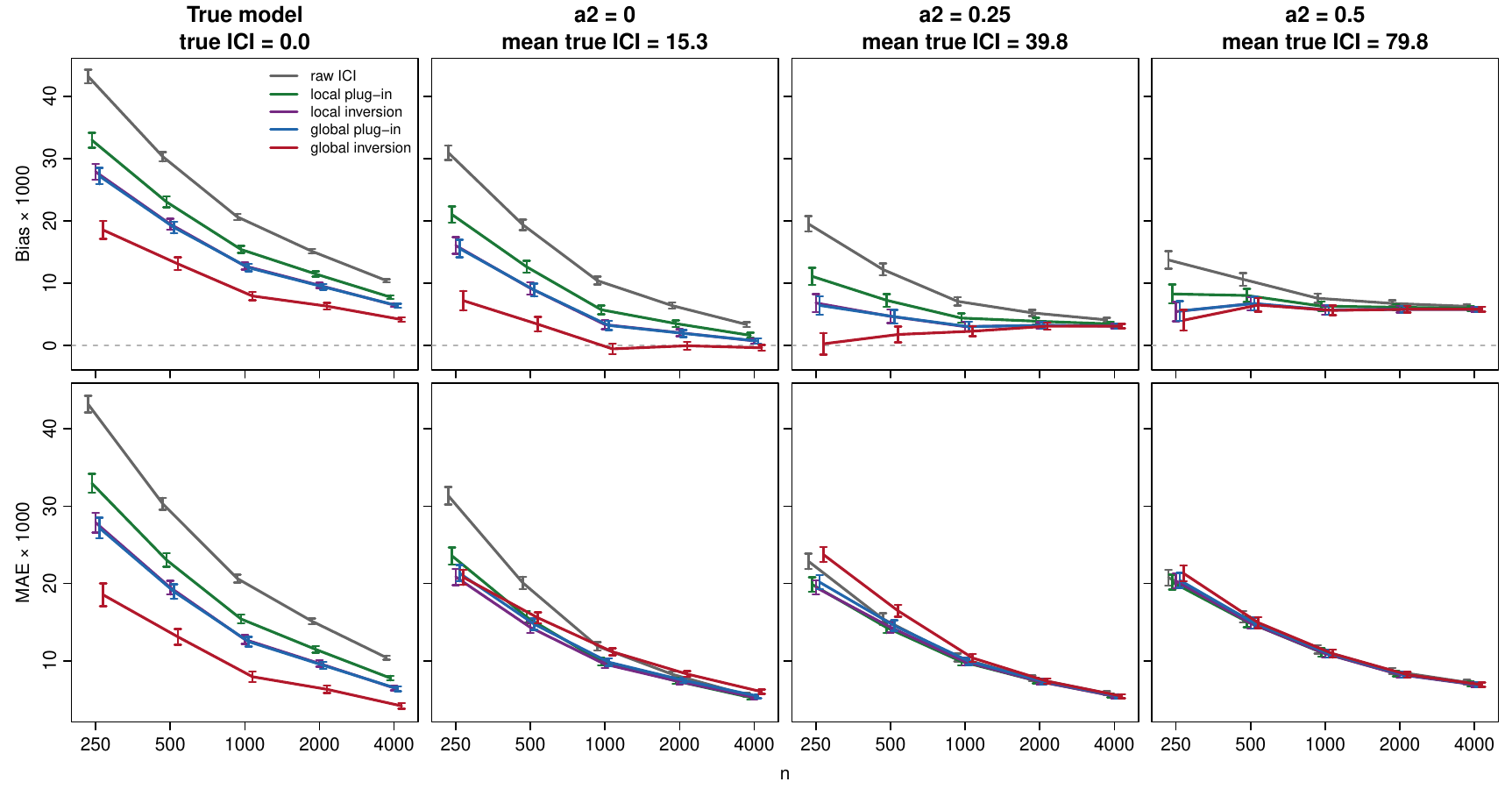}
\par\smallskip
{\footnotesize\raggedright \textit{Note:} Layout and conventions as in Figure
\ref{fig:simquad}. ICI: integrated calibration index; MAE: mean absolute error.\par}
\end{figure}

\begin{figure}[H]
\centering
\caption{Simulation results, restricted cubic spline, interaction design.}
\label{fig:rcsinter}
\includegraphics[width=0.9\textwidth]{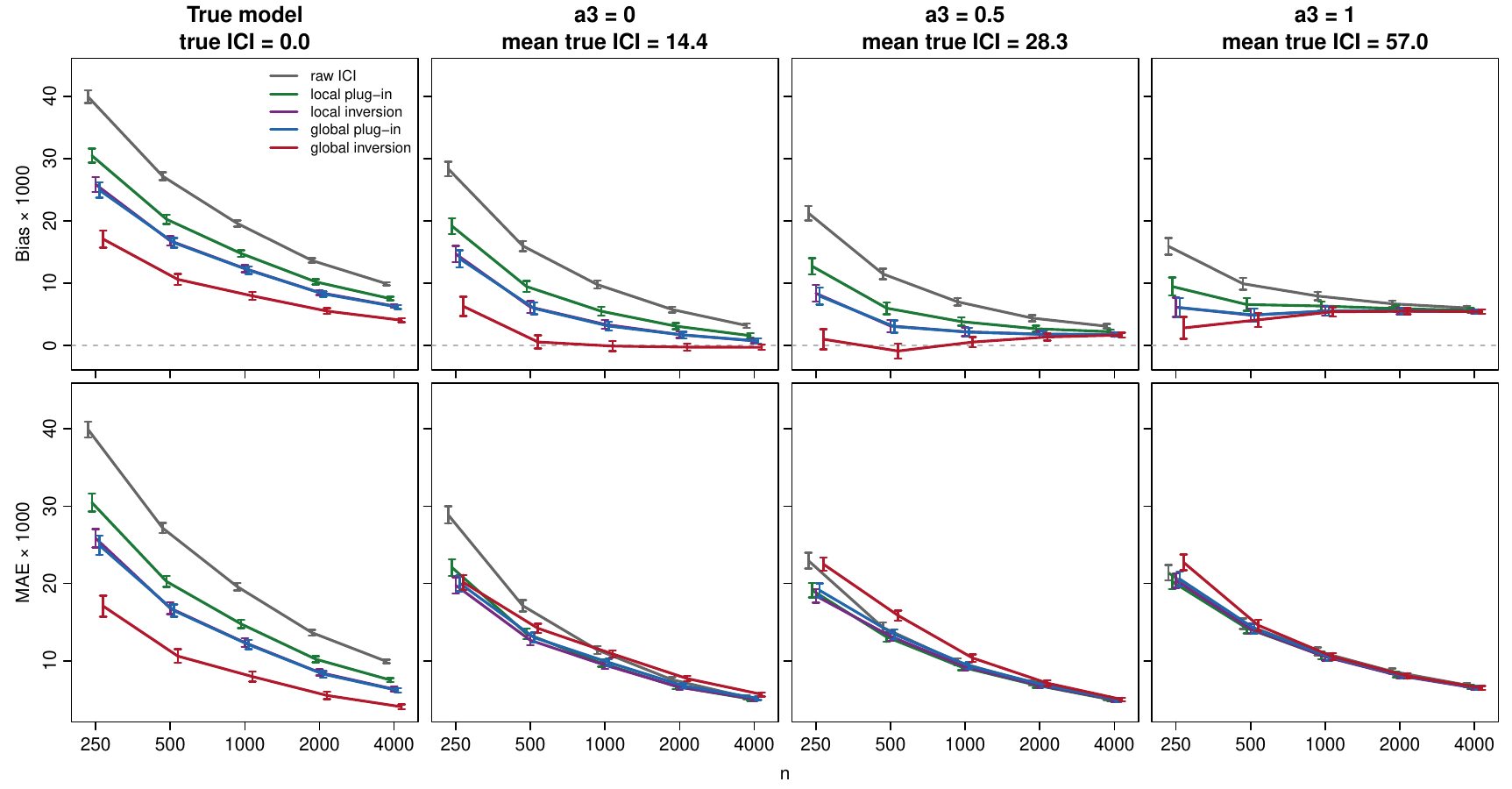}
\par\smallskip
{\footnotesize\raggedright \textit{Note:} Layout and conventions as in Figure
\ref{fig:simquad}. ICI: integrated calibration index; MAE: mean absolute error.\par}
\end{figure}

\begin{figure}[H]
\centering
\caption{Simulation results, generalized additive model, quadratic design.}
\label{fig:gamquad}
\includegraphics[width=0.9\textwidth]{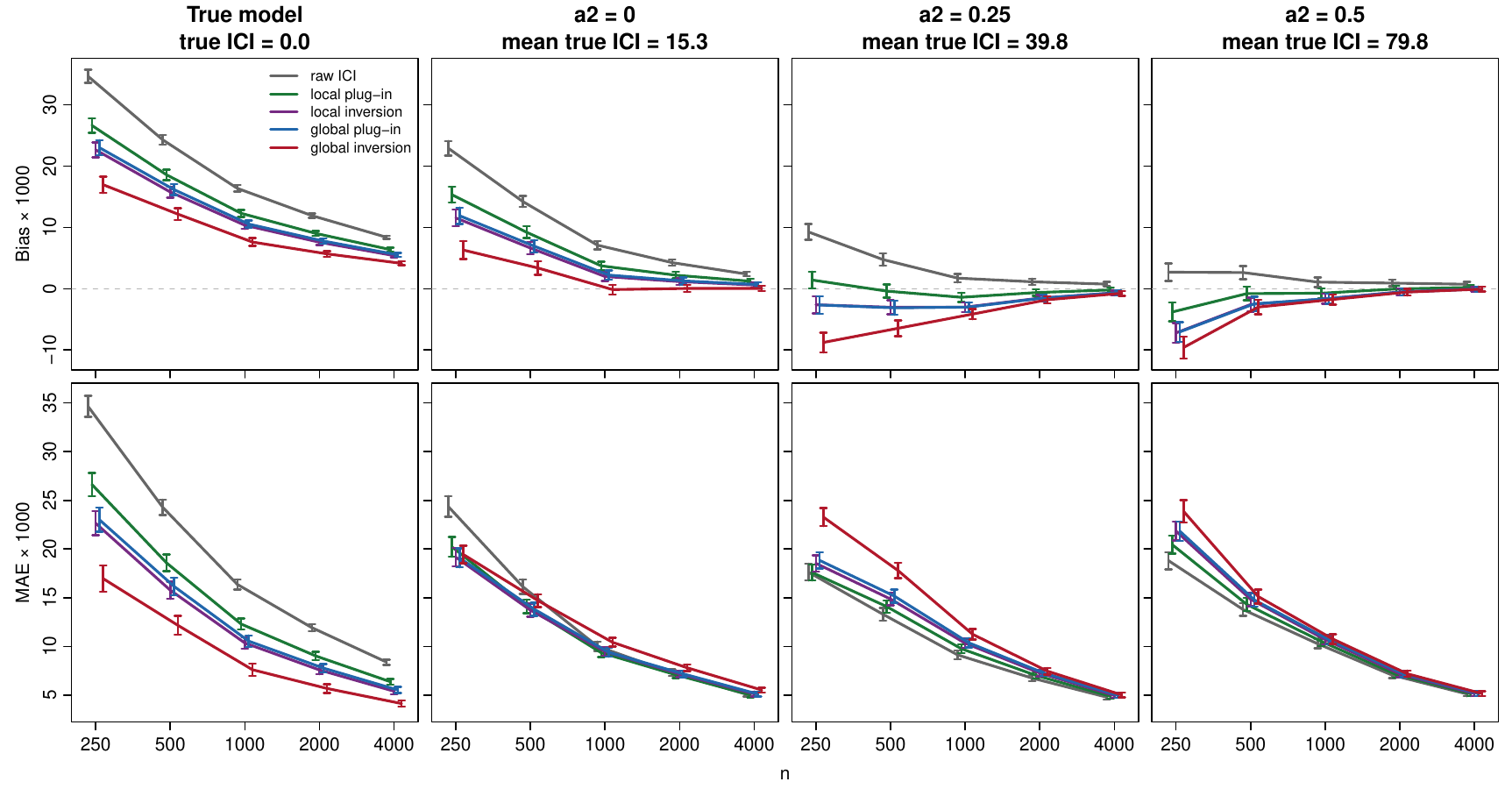}
\par\smallskip
{\footnotesize\raggedright \textit{Note:} Layout and conventions as in Figure
\ref{fig:simquad}. ICI: integrated calibration index; MAE: mean absolute error.\par}
\end{figure}

\begin{figure}[H]
\centering
\caption{Simulation results, generalized additive model, interaction design.}
\label{fig:gaminter}
\includegraphics[width=0.9\textwidth]{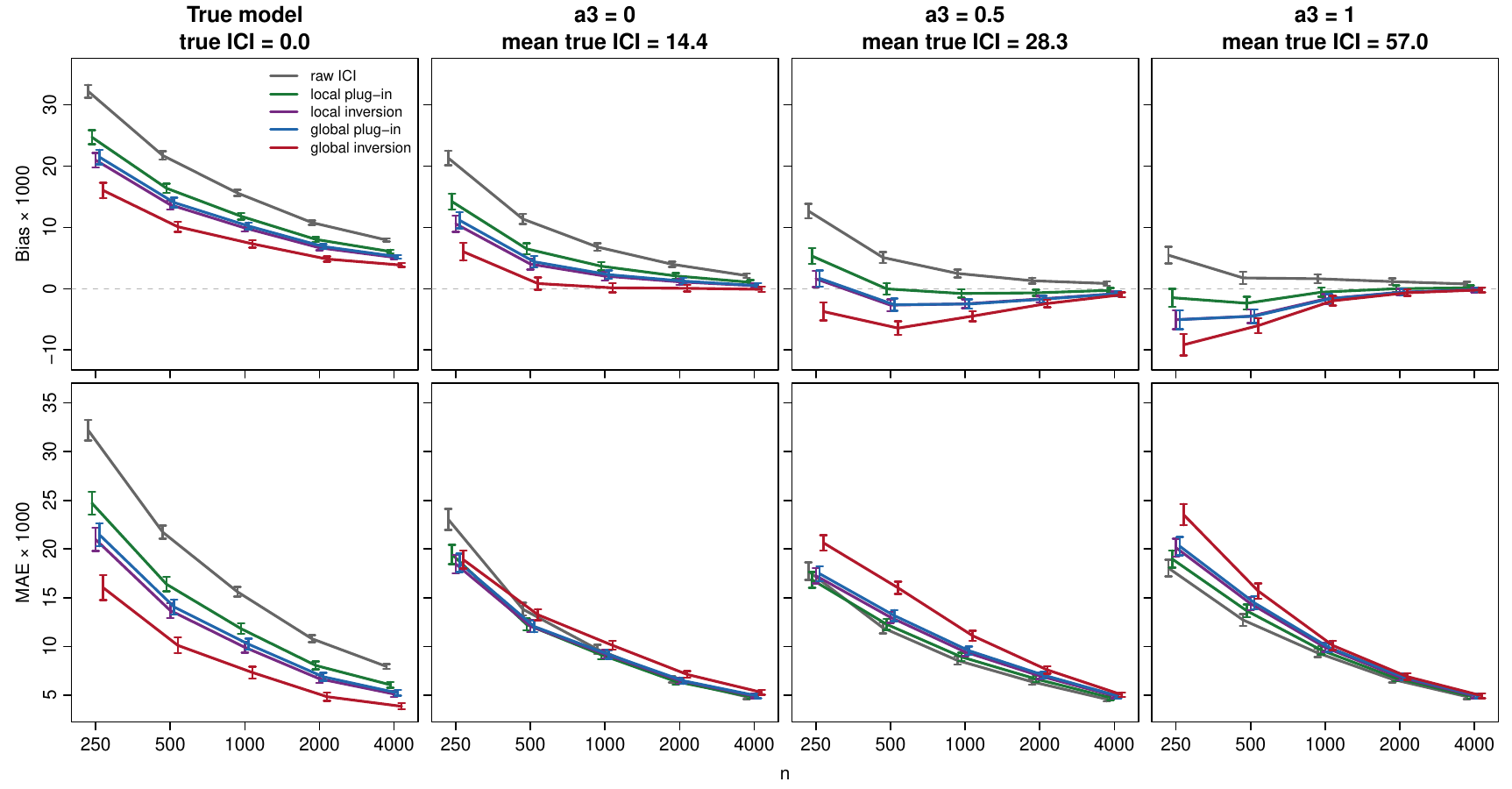}
\par\smallskip
{\footnotesize\raggedright \textit{Note:} Layout and conventions as in Figure
\ref{fig:simquad}. ICI: integrated calibration index; MAE: mean absolute error.\par}
\end{figure}

\clearpage

\section{Quantile-based summaries (E50 and E90)}\label{app-quantiles}

\begin{figure}[H]
\centering
\caption{E50 read off the adjusted error curves, quadratic design (LOESS).}
\label{fig:e50quad}
\includegraphics[width=0.9\textwidth]{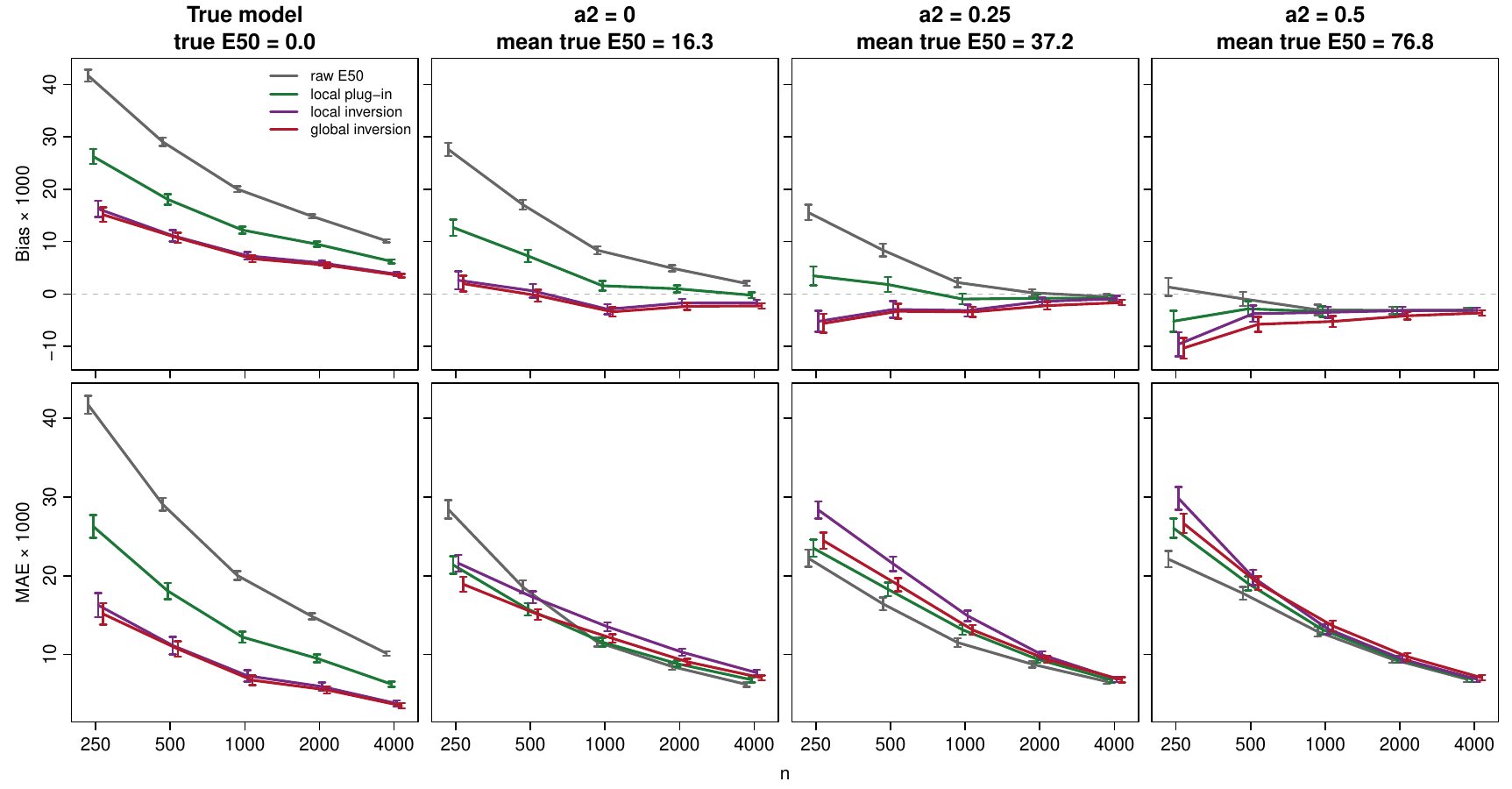}
\par\smallskip
{\footnotesize\raggedright \textit{Note:} Layout and conventions as in Figure
\ref{fig:simquad}; GP is not shown, as its read-offs equal those of LP. E50, E90:
median and 90th percentile of the absolute calibration error; MAE: mean absolute
error; LOESS: locally estimated scatterplot smoothing.\par}
\end{figure}

\begin{figure}[H]
\centering
\caption{E50 read off the adjusted error curves, interaction design (LOESS).}
\label{fig:e50inter}
\includegraphics[width=0.9\textwidth]{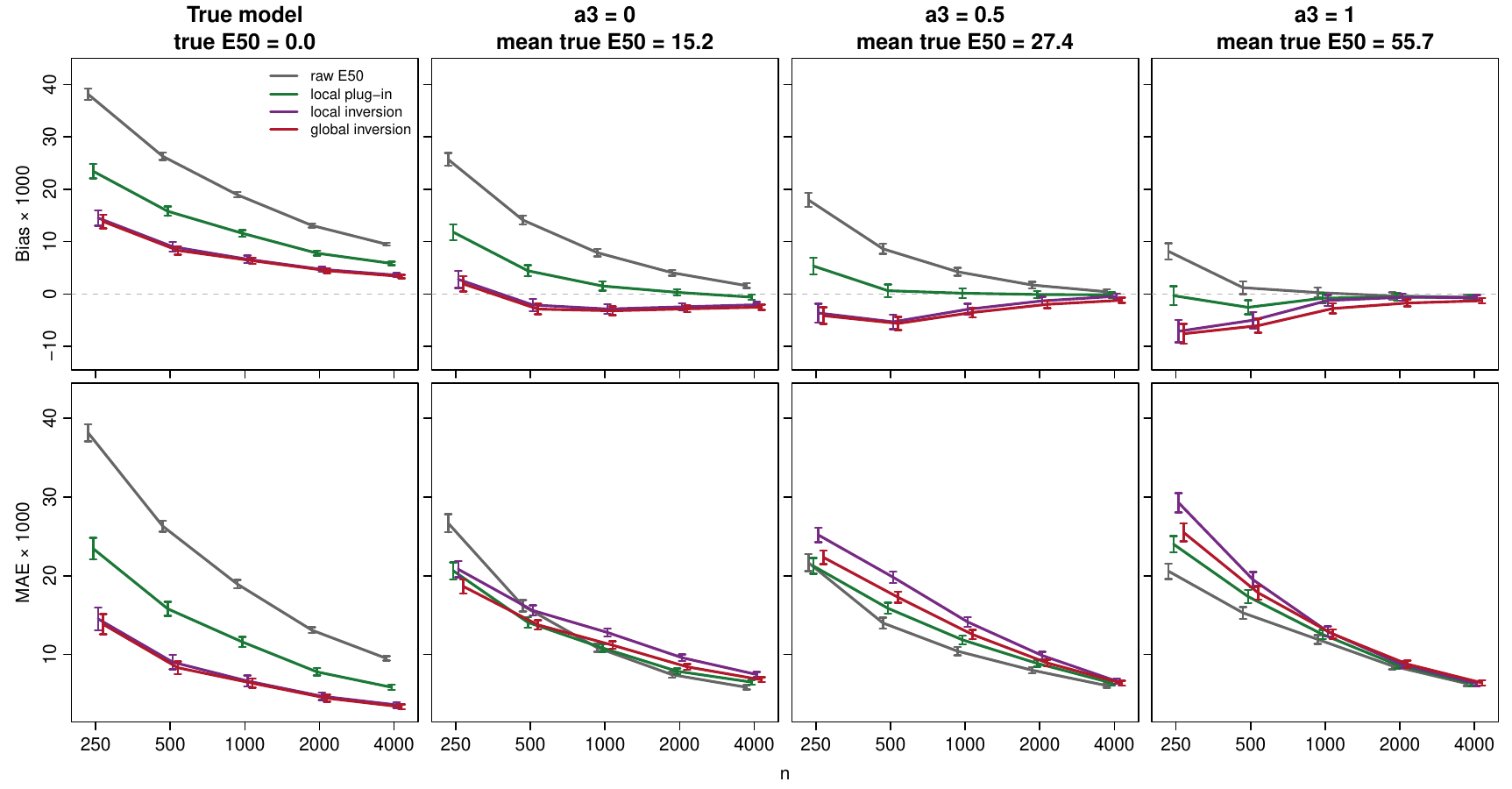}
\par\smallskip
{\footnotesize\raggedright \textit{Note:} Layout and conventions as in Figure
\ref{fig:simquad}; GP is not shown, as its read-offs equal those of LP. E50, E90:
median and 90th percentile of the absolute calibration error; MAE: mean absolute
error; LOESS: locally estimated scatterplot smoothing.\par}
\end{figure}

\begin{figure}[H]
\centering
\caption{E90 read off the adjusted error curves, quadratic design (LOESS).}
\label{fig:e90quad}
\includegraphics[width=0.9\textwidth]{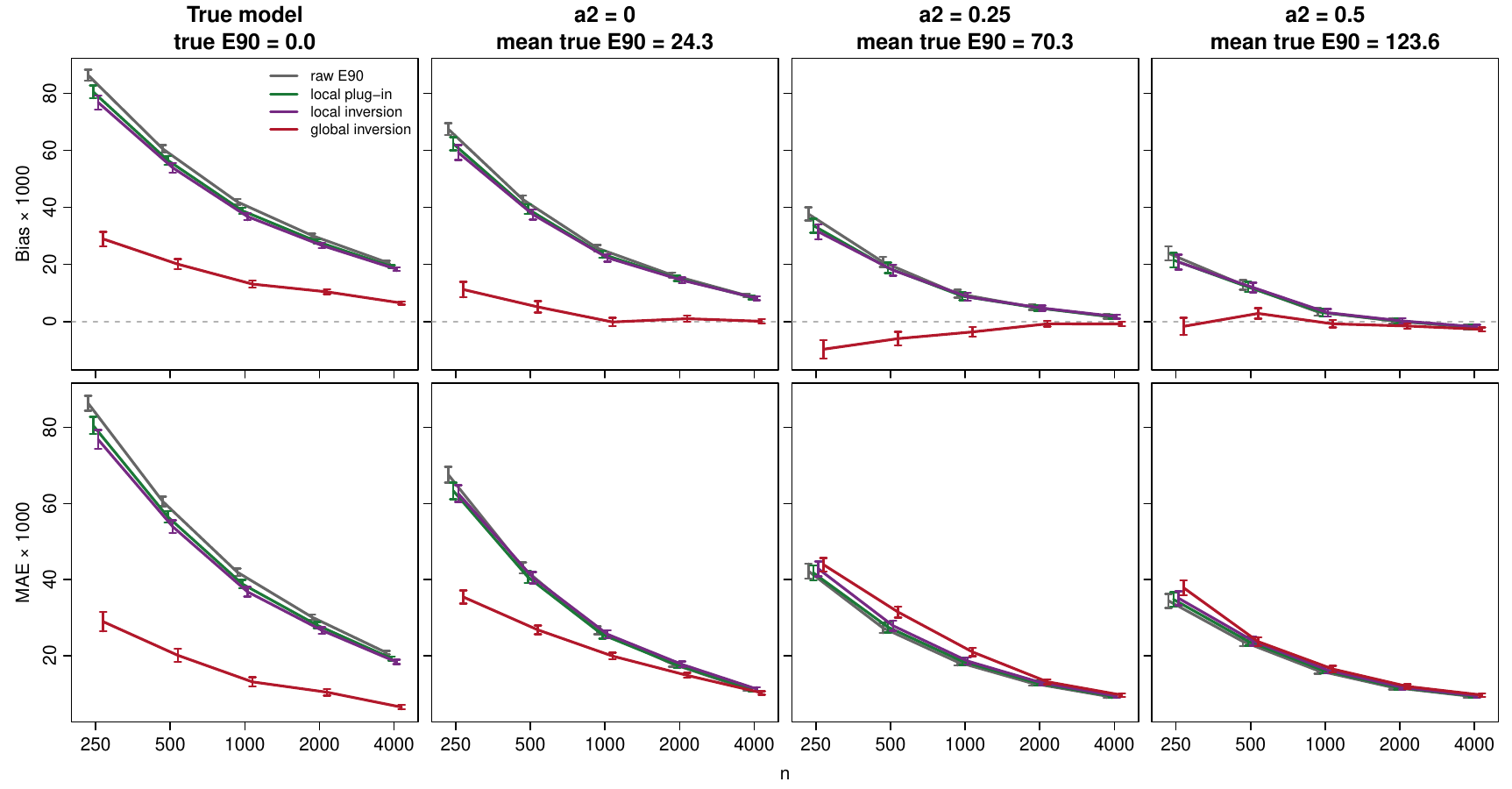}
\par\smallskip
{\footnotesize\raggedright \textit{Note:} Layout and conventions as in Figure
\ref{fig:simquad}; GP is not shown, as its read-offs equal those of LP. E50, E90:
median and 90th percentile of the absolute calibration error; MAE: mean absolute
error; LOESS: locally estimated scatterplot smoothing.\par}
\end{figure}

\begin{figure}[H]
\centering
\caption{E90 read off the adjusted error curves, interaction design (LOESS).}
\label{fig:e90inter}
\includegraphics[width=0.9\textwidth]{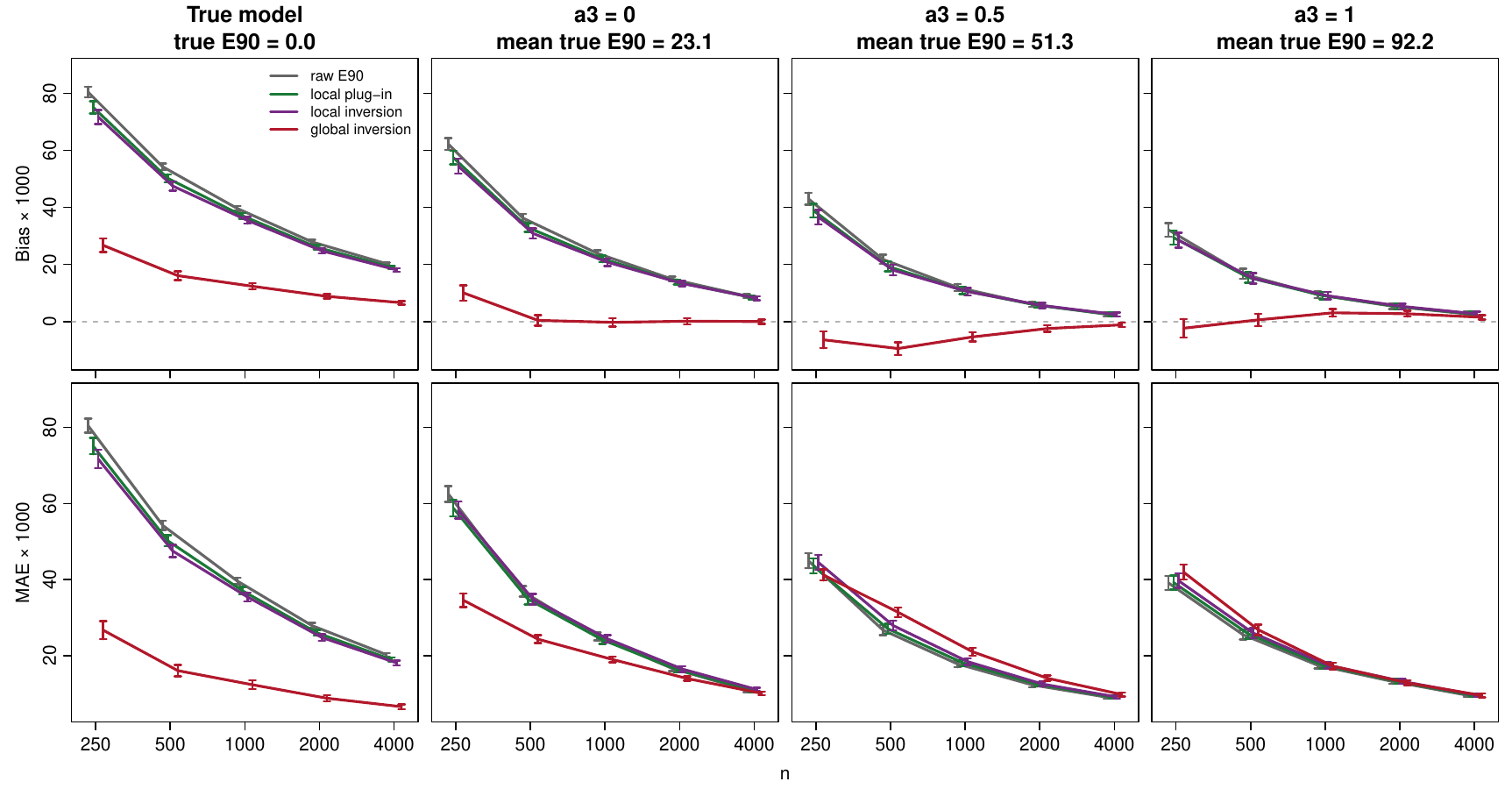}
\par\smallskip
{\footnotesize\raggedright \textit{Note:} Layout and conventions as in Figure
\ref{fig:simquad}; GP is not shown, as its read-offs equal those of LP. E50, E90:
median and 90th percentile of the absolute calibration error; MAE: mean absolute
error; LOESS: locally estimated scatterplot smoothing.\par}
\end{figure}

\begin{figure}[H]
\centering
\caption{E50 read off the adjusted error curves, quadratic design (restricted cubic spline).}
\label{fig:e50rcsquad}
\includegraphics[width=0.9\textwidth]{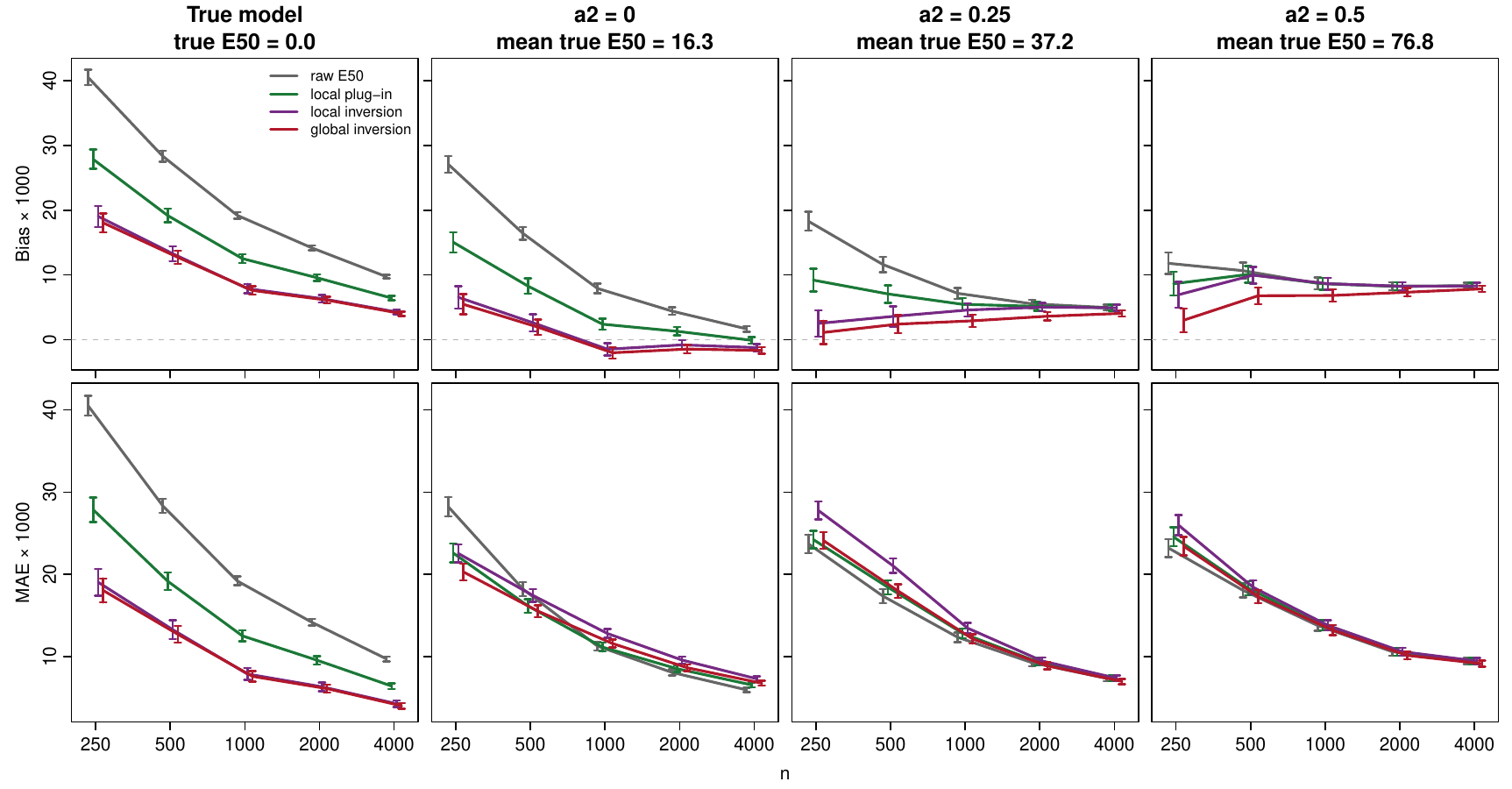}
\par\smallskip
{\footnotesize\raggedright \textit{Note:} Layout and conventions as in Figure
\ref{fig:simquad}; GP is not shown, as its read-offs equal those of LP. E50, E90:
median and 90th percentile of the absolute calibration error; MAE: mean absolute
error.\par}
\end{figure}

\begin{figure}[H]
\centering
\caption{E50 read off the adjusted error curves, interaction design (restricted cubic spline).}
\label{fig:e50rcsinter}
\includegraphics[width=0.9\textwidth]{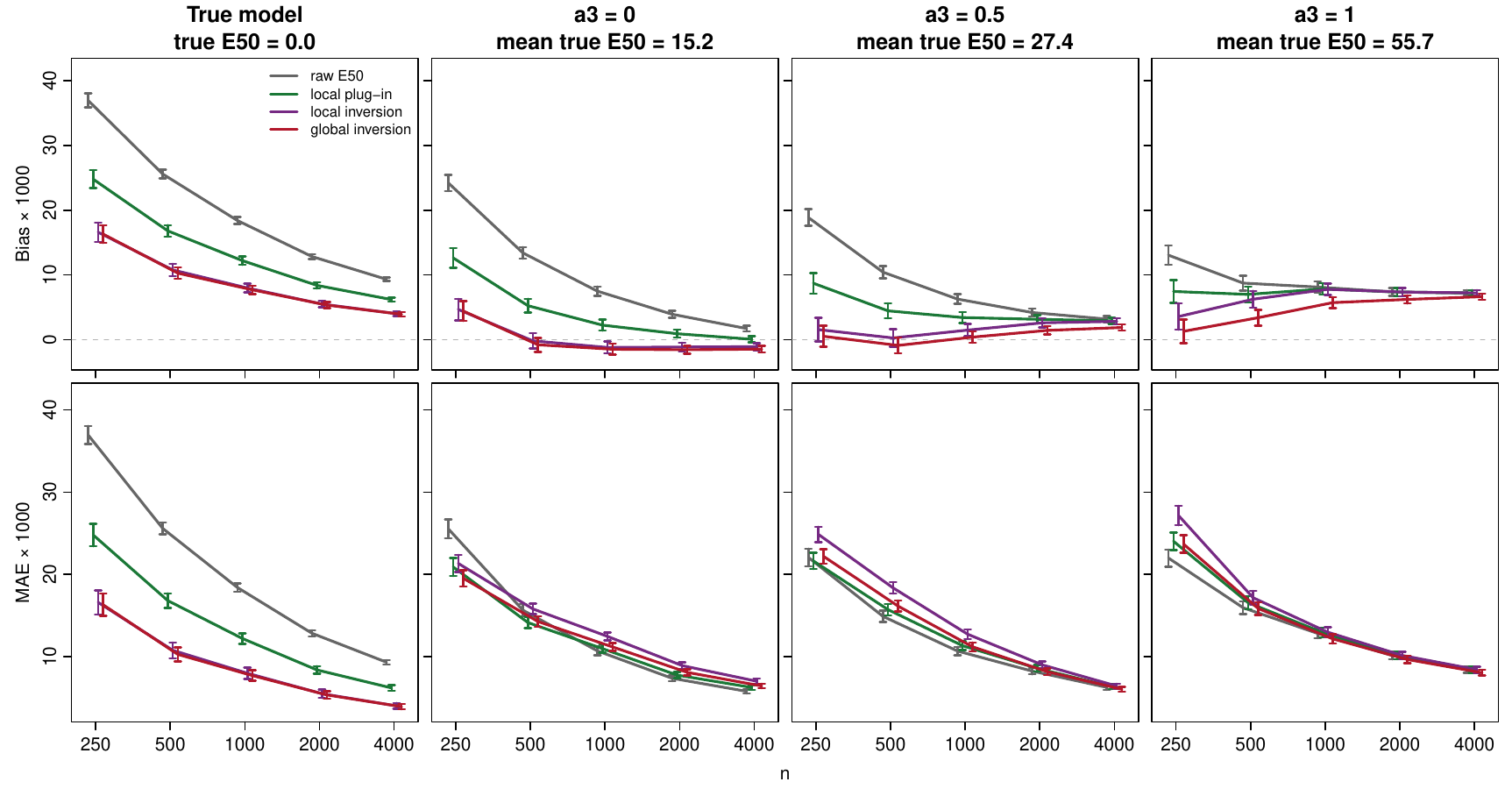}
\par\smallskip
{\footnotesize\raggedright \textit{Note:} Layout and conventions as in Figure
\ref{fig:simquad}; GP is not shown, as its read-offs equal those of LP. E50, E90:
median and 90th percentile of the absolute calibration error; MAE: mean absolute
error.\par}
\end{figure}

\begin{figure}[H]
\centering
\caption{E90 read off the adjusted error curves, quadratic design (restricted cubic spline).}
\label{fig:e90rcsquad}
\includegraphics[width=0.9\textwidth]{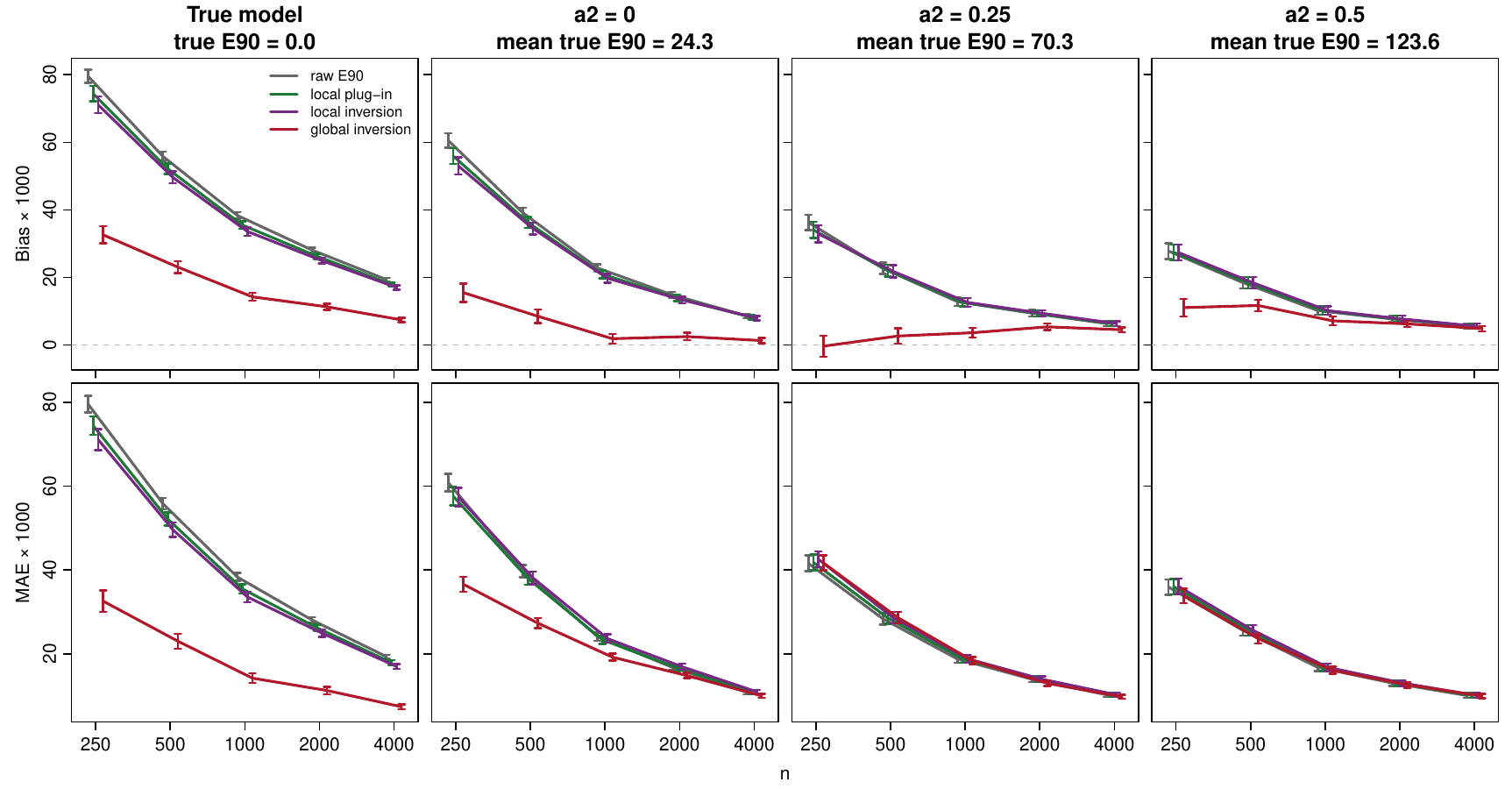}
\par\smallskip
{\footnotesize\raggedright \textit{Note:} Layout and conventions as in Figure
\ref{fig:simquad}; GP is not shown, as its read-offs equal those of LP. E50, E90:
median and 90th percentile of the absolute calibration error; MAE: mean absolute
error.\par}
\end{figure}

\begin{figure}[H]
\centering
\caption{E90 read off the adjusted error curves, interaction design (restricted cubic spline).}
\label{fig:e90rcsinter}
\includegraphics[width=0.9\textwidth]{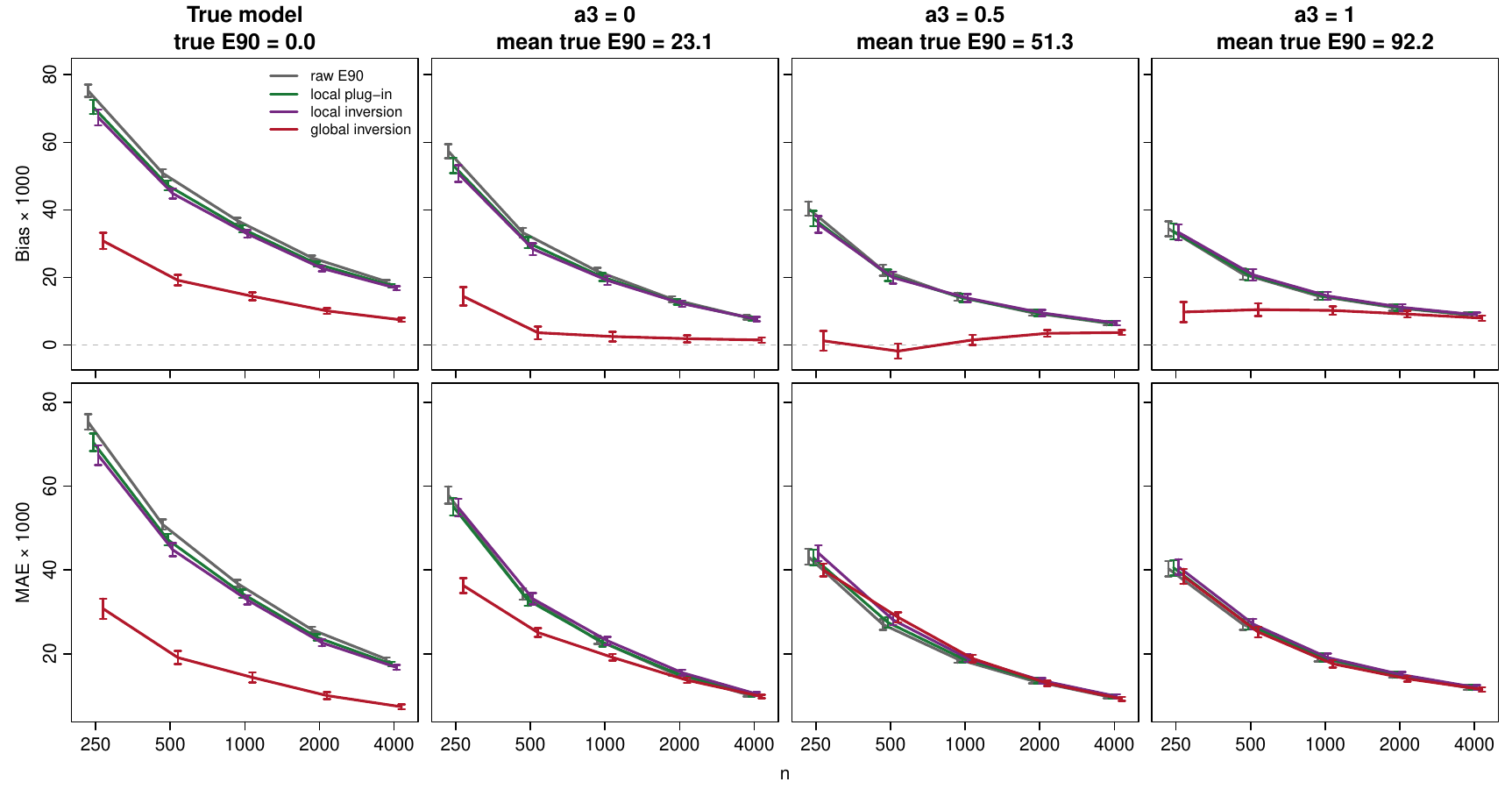}
\par\smallskip
{\footnotesize\raggedright \textit{Note:} Layout and conventions as in Figure
\ref{fig:simquad}; GP is not shown, as its read-offs equal those of LP. E50, E90:
median and 90th percentile of the absolute calibration error; MAE: mean absolute
error.\par}
\end{figure}

\begin{figure}[H]
\centering
\caption{E50 read off the adjusted error curves, quadratic design (generalized additive model).}
\label{fig:e50gamquad}
\includegraphics[width=0.9\textwidth]{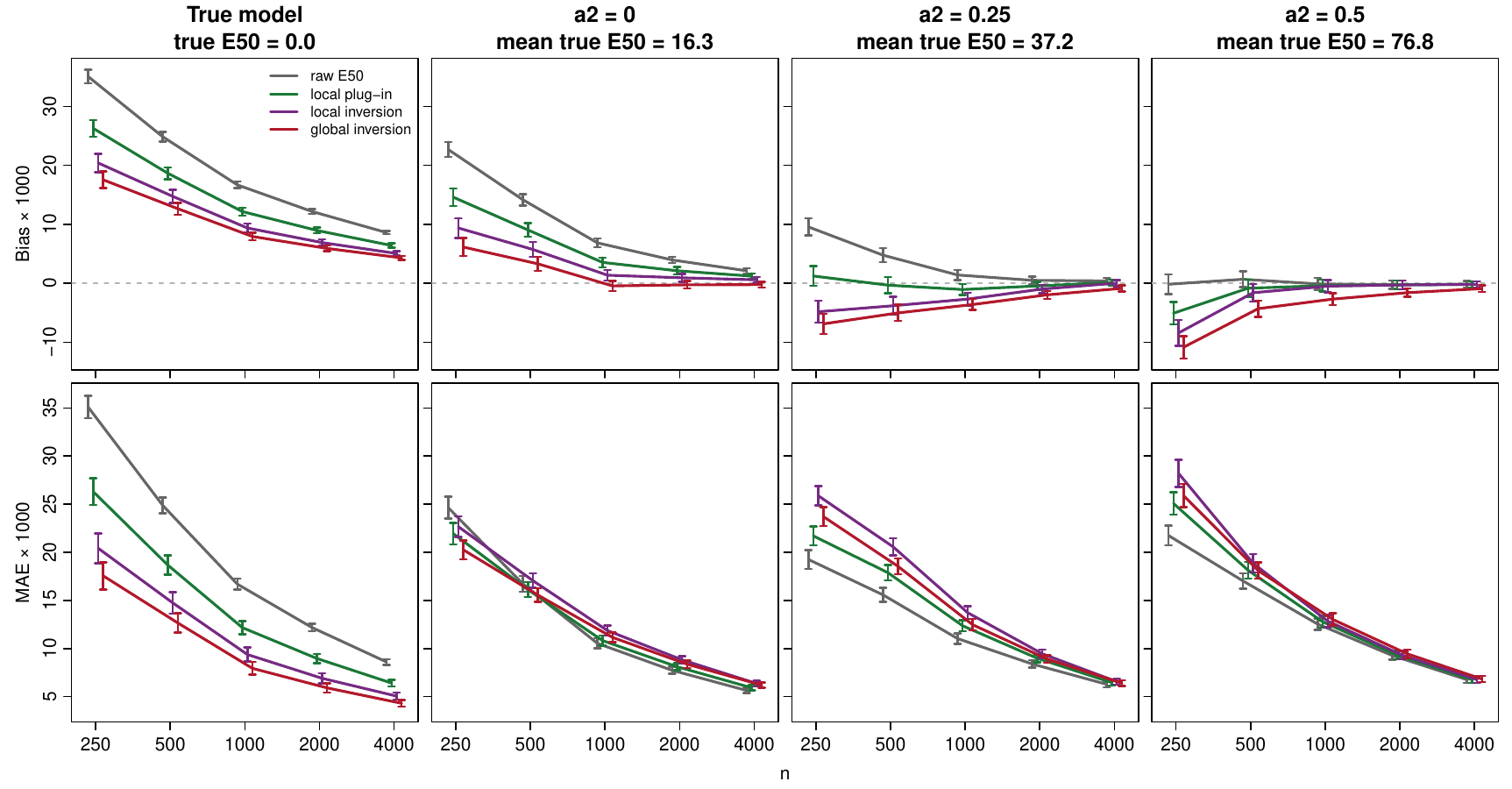}
\par\smallskip
{\footnotesize\raggedright \textit{Note:} Layout and conventions as in Figure
\ref{fig:simquad}; GP is not shown, as its read-offs equal those of LP. E50, E90:
median and 90th percentile of the absolute calibration error; MAE: mean absolute
error.\par}
\end{figure}

\begin{figure}[H]
\centering
\caption{E50 read off the adjusted error curves, interaction design (generalized additive model).}
\label{fig:e50gaminter}
\includegraphics[width=0.9\textwidth]{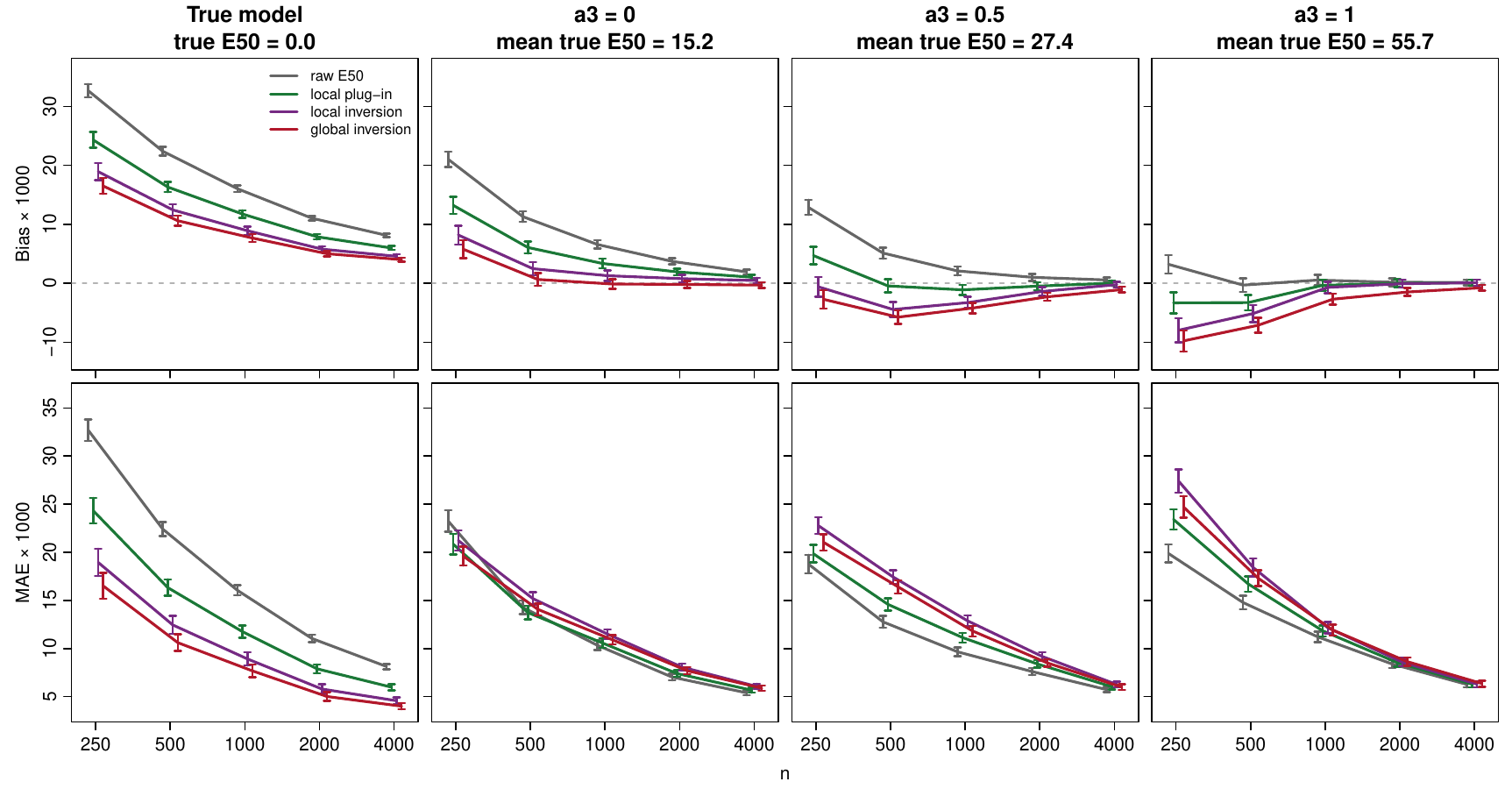}
\par\smallskip
{\footnotesize\raggedright \textit{Note:} Layout and conventions as in Figure
\ref{fig:simquad}; GP is not shown, as its read-offs equal those of LP. E50, E90:
median and 90th percentile of the absolute calibration error; MAE: mean absolute
error.\par}
\end{figure}

\begin{figure}[H]
\centering
\caption{E90 read off the adjusted error curves, quadratic design (generalized additive model).}
\label{fig:e90gamquad}
\includegraphics[width=0.9\textwidth]{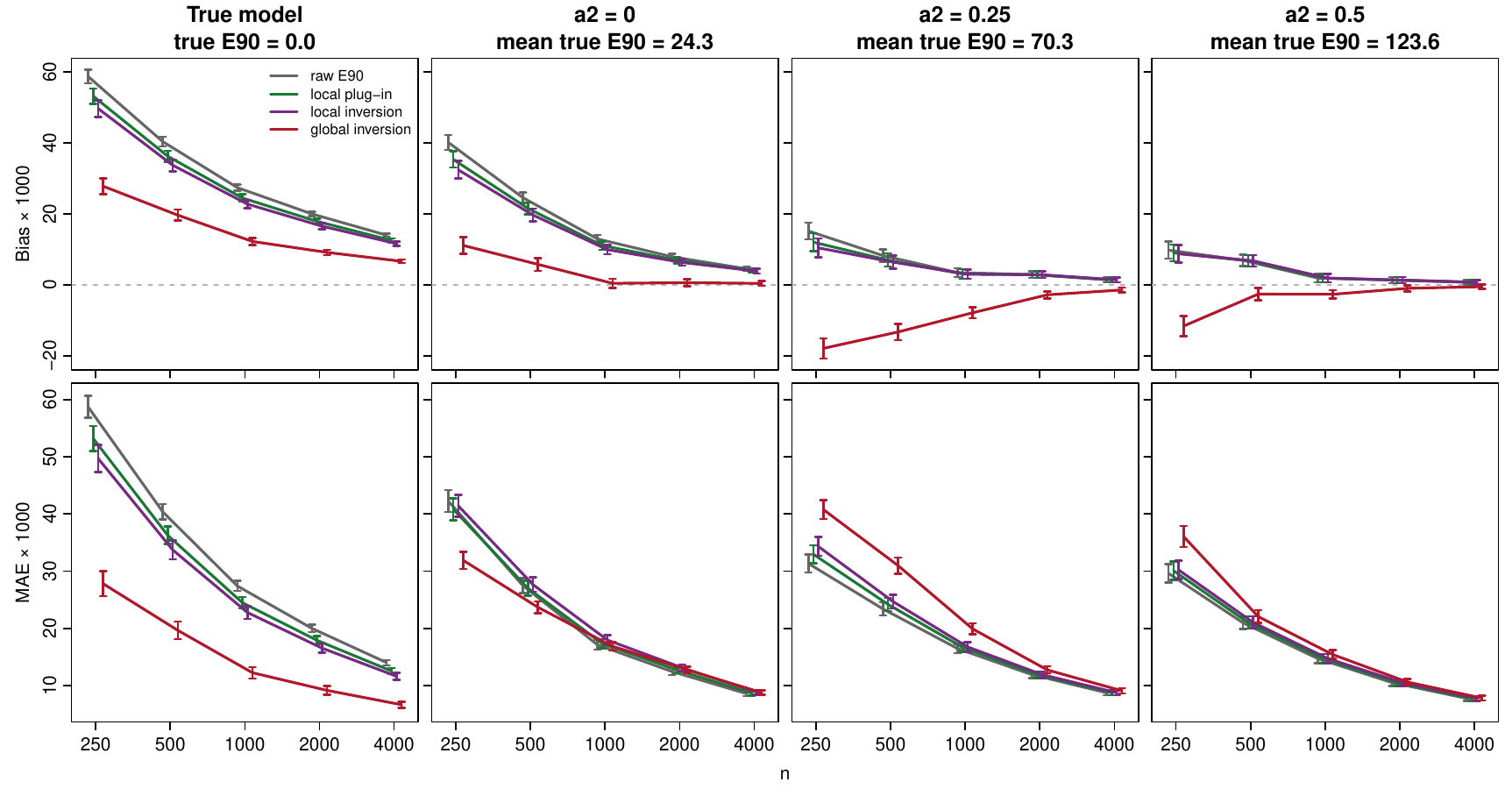}
\par\smallskip
{\footnotesize\raggedright \textit{Note:} Layout and conventions as in Figure
\ref{fig:simquad}; GP is not shown, as its read-offs equal those of LP. E50, E90:
median and 90th percentile of the absolute calibration error; MAE: mean absolute
error.\par}
\end{figure}

\begin{figure}[H]
\centering
\caption{E90 read off the adjusted error curves, interaction design (generalized additive model).}
\label{fig:e90gaminter}
\includegraphics[width=0.9\textwidth]{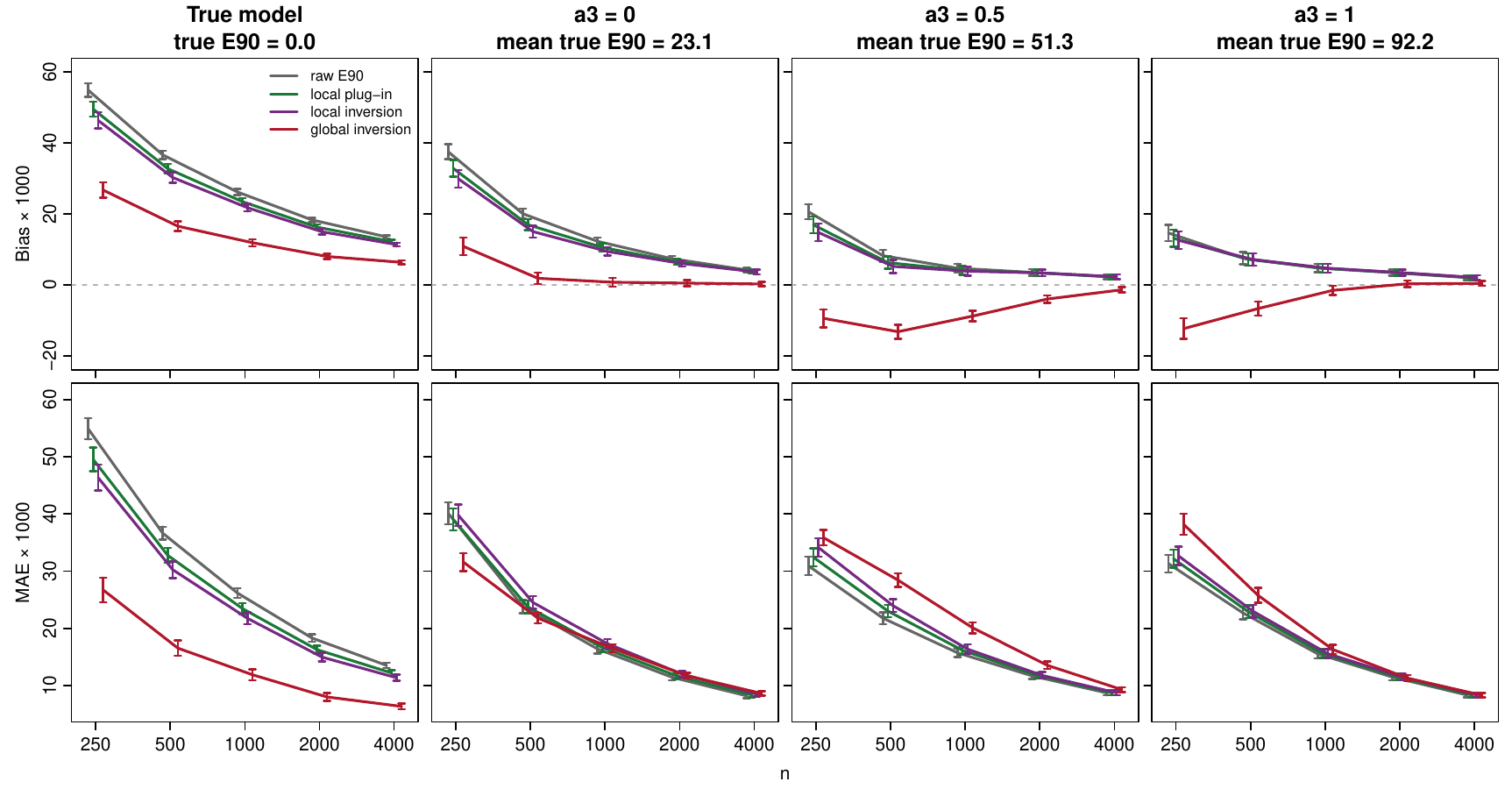}
\par\smallskip
{\footnotesize\raggedright \textit{Note:} Layout and conventions as in Figure
\ref{fig:simquad}; GP is not shown, as its read-offs equal those of LP. E50, E90:
median and 90th percentile of the absolute calibration error; MAE: mean absolute
error.\par}
\end{figure}

We evaluated the performance of the proposed methods for mitigating the
bias of quantile-based summary indices of miscalibration (E50: median
absolute calibration error, E90: 90th percentile absolute calibration
error). These metrics were assessed in the main simulation loop
alongside ICI, using the same three smoothers: LOESS in Figures
\ref{fig:e50quad} to \ref{fig:e90inter}, the restricted cubic spline in
Figures \ref{fig:e50rcsquad} to \ref{fig:e90rcsinter}, and the
generalized additive model in Figures \ref{fig:e50gamquad} to
\ref{fig:e90gaminter}. The true E50 and E90 of each model are the median
and 90th percentile of true absolute calibration errors, computed in the
same way as the true ICI. The layout and units follow those of Figures
\ref{fig:simquad} and \ref{fig:siminter}. Across the three smoothers,
the Monte Carlo standard error of a difference from the raw summary is
at most 0.40 (E50) and 0.79 (E90) for bias, and 0.59 and 1.22 for MAE.

\clearpage

\section{Reference implementation}\label{app-code}

The GI estimator needs only a point estimate of the calibration error at
each observation together with its standard error, so it applies to any
smoother that reports both. The function below is self-contained and
uses nothing beyond base R (\texttt{uniroot} and \texttt{pnorm}, both in
the default \texttt{stats} package).

\begin{Shaded}
\begin{Highlighting}[]
\DocumentationTok{\#\# Bias{-}mitigated ICI by global inversion (GI).}
\DocumentationTok{\#\#   ehat : smoothed calibration error at each observation, ghat(p\_i) {-} p\_i}
\DocumentationTok{\#\#   se   : standard error of the smoother at the same points}
\DocumentationTok{\#\# Returns the corrected ICI, the multiplier k, the raw ICI and the noise floor.}
\NormalTok{ici\_gi }\OtherTok{\textless{}{-}} \ControlFlowTok{function}\NormalTok{(ehat, se) \{}
  \ControlFlowTok{if}\NormalTok{ (}\FunctionTok{any}\NormalTok{(}\SpecialCharTok{!}\FunctionTok{is.finite}\NormalTok{(ehat) }\SpecialCharTok{|} \SpecialCharTok{!}\FunctionTok{is.finite}\NormalTok{(se) }\SpecialCharTok{|}\NormalTok{ se }\SpecialCharTok{\textless{}=} \DecValTok{0}\NormalTok{)) }
    \FunctionTok{stop}\NormalTok{(}\StringTok{"Inadmissible points."}\NormalTok{)}
  
  \DocumentationTok{\#\# mean of |X| for X \textasciitilde{} N(mu, sigma\^{}2): the folded normal mean}
\NormalTok{  mu\_fn }\OtherTok{\textless{}{-}} \ControlFlowTok{function}\NormalTok{(mu, sigma) \{}
\NormalTok{    m }\OtherTok{\textless{}{-}}\NormalTok{ mu }\SpecialCharTok{/}\NormalTok{ sigma}
\NormalTok{    sigma }\SpecialCharTok{*}\NormalTok{ (}\FunctionTok{sqrt}\NormalTok{(}\DecValTok{2} \SpecialCharTok{/}\NormalTok{ pi) }\SpecialCharTok{*} \FunctionTok{exp}\NormalTok{(}\SpecialCharTok{{-}}\NormalTok{m}\SpecialCharTok{\^{}}\DecValTok{2} \SpecialCharTok{/} \DecValTok{2}\NormalTok{) }\SpecialCharTok{+}\NormalTok{ m }\SpecialCharTok{*}\NormalTok{ (}\DecValTok{2} \SpecialCharTok{*} \FunctionTok{pnorm}\NormalTok{(m) }\SpecialCharTok{{-}} \DecValTok{1}\NormalTok{))}
\NormalTok{  \}}

\NormalTok{  ici    }\OtherTok{\textless{}{-}} \FunctionTok{mean}\NormalTok{(}\FunctionTok{abs}\NormalTok{(ehat))                        }
\NormalTok{  theta  }\OtherTok{\textless{}{-}} \ControlFlowTok{function}\NormalTok{(k) }\FunctionTok{mean}\NormalTok{(}\FunctionTok{mu\_fn}\NormalTok{(k }\SpecialCharTok{*}\NormalTok{ ehat, se))   }
\NormalTok{  nfloor }\OtherTok{\textless{}{-}} \FunctionTok{theta}\NormalTok{(}\DecValTok{0}\NormalTok{)                               }

  \DocumentationTok{\#\# theta is non{-}decreasing in k, with theta(0) = nfloor and theta(1) \textgreater{}= ici,}
  \DocumentationTok{\#\# so a root always exists in [0, 1]; it is 0 exactly when ici \textless{}= nfloor.}
\NormalTok{  k }\OtherTok{\textless{}{-}} \ControlFlowTok{if}\NormalTok{ (nfloor }\SpecialCharTok{\textgreater{}=}\NormalTok{ ici) }\DecValTok{0} \ControlFlowTok{else}
       \FunctionTok{uniroot}\NormalTok{(}\ControlFlowTok{function}\NormalTok{(k) }\FunctionTok{theta}\NormalTok{(k) }\SpecialCharTok{{-}}\NormalTok{ ici, }\FunctionTok{c}\NormalTok{(}\DecValTok{0}\NormalTok{, }\DecValTok{1}\NormalTok{), }\AttributeTok{tol =} \FloatTok{1e{-}10}\NormalTok{)}\SpecialCharTok{$}\NormalTok{root}

  \FunctionTok{c}\NormalTok{(}\AttributeTok{ici\_gi =}\NormalTok{ k }\SpecialCharTok{*}\NormalTok{ ici, }\AttributeTok{k =}\NormalTok{ k, }\AttributeTok{ici\_raw =}\NormalTok{ ici, }\AttributeTok{floor =}\NormalTok{ nfloor)}
\NormalTok{\}}
\end{Highlighting}
\end{Shaded}

A worked example using \texttt{infert}, a 248-subject dataset with a
binary outcome that ships with base R. The predicted risks come from a
model with fixed, pre-specified coefficients, so no data are spent on
fitting and the whole sample is available for the calibration
assessment:

\begin{Shaded}
\begin{Highlighting}[]
\NormalTok{b }\OtherTok{\textless{}{-}} \FunctionTok{c}\NormalTok{(}\SpecialCharTok{{-}}\FloatTok{3.0}\NormalTok{, }\FloatTok{0.05}\NormalTok{, }\SpecialCharTok{{-}}\FloatTok{0.7}\NormalTok{, }\FloatTok{1.2}\NormalTok{, }\FloatTok{1.9}\NormalTok{) }\CommentTok{\# intercept,age,parity,induced,spontaneous}
\NormalTok{X }\OtherTok{\textless{}{-}} \FunctionTok{cbind}\NormalTok{(}\DecValTok{1}\NormalTok{, infert}\SpecialCharTok{$}\NormalTok{age, infert}\SpecialCharTok{$}\NormalTok{parity, infert}\SpecialCharTok{$}\NormalTok{induced, infert}\SpecialCharTok{$}\NormalTok{spontaneous)}
\NormalTok{p }\OtherTok{\textless{}{-}} \FunctionTok{plogis}\NormalTok{(}\FunctionTok{as.vector}\NormalTok{(X }\SpecialCharTok{\%*\%}\NormalTok{ b))}
\NormalTok{y }\OtherTok{\textless{}{-}}\NormalTok{ infert}\SpecialCharTok{$}\NormalTok{case}

\NormalTok{fit }\OtherTok{\textless{}{-}} \FunctionTok{loess}\NormalTok{(y }\SpecialCharTok{\textasciitilde{}}\NormalTok{ p, }\AttributeTok{degree =} \DecValTok{2}\NormalTok{, }\AttributeTok{span =} \FloatTok{0.75}\NormalTok{)}
\NormalTok{g   }\OtherTok{\textless{}{-}} \FunctionTok{predict}\NormalTok{(fit, }\AttributeTok{se =} \ConstantTok{TRUE}\NormalTok{)}
\FunctionTok{ici\_gi}\NormalTok{(g}\SpecialCharTok{$}\NormalTok{fit }\SpecialCharTok{{-}}\NormalTok{ p, g}\SpecialCharTok{$}\NormalTok{se.fit)}
\end{Highlighting}
\end{Shaded}

\begin{verbatim}
    ici_gi          k    ici_raw      floor 
0.03761229 0.67770000 0.05549991 0.04330491 
\end{verbatim}

The raw ICI of 0.055 is only modestly above the noise floor of 0.043
that a perfectly calibrated model would produce at this sample size. The
correction rescales the fitted calibration error by \(k = 0.68\) and
returns 0.038, about a third lower than the uncorrected value.

\end{document}